\documentclass[aps,prc,twocolumn,showpacs,floatfix,superscriptaddress,nofootinbib,longbibliography]{revtex4-1}
\usepackage[utf8]{inputenc}

\usepackage{amsmath}
\usepackage{amssymb}
\usepackage{inputenc}
\usepackage{amssymb}
\usepackage{graphicx}
\usepackage{epsfig}
\usepackage{bm}
\usepackage{color}
\usepackage{float}
\usepackage{dcolumn}
\usepackage{multirow} 
\usepackage{MnSymbol}
\usepackage[unicode=true,
  linktocpage,
  linkbordercolor={0.5 0.5 1},
  citebordercolor={0.5 1 0.5},
  colorlinks=true,
  linkcolor=blue,
  citecolor=blue,
  urlcolor=blue]{hyperref} 
\usepackage{lipsum} 
\usepackage{placeins} 
\usepackage[dvipsnames,usenames]{xcolor}

\begin{document}
\title{$^{16}$O spectral function  from coupled-cluster theory:\\ applications to lepton-nucleus scattering}

\date{\today}
\author{J.~E.~Sobczyk}
\affiliation{Institut f\"ur Kernphysik and PRISMA$^+$ Cluster of Excellence, Johannes Gutenberg-Universit\"at, 55128
  Mainz, Germany}
\author{S.~Bacca}
\affiliation{Institut f\"ur Kernphysik and PRISMA$^+$ Cluster of Excellence, Johannes Gutenberg-Universit\"at, 55128
  Mainz, Germany}
  \affiliation{Helmholtz-Institut Mainz, Johannes Gutenberg-Universit\"at Mainz, D-55099 Mainz, Germany}

\begin{abstract}
We calculate the $^{16}$O  spectral function by combining coupled-cluster theory with a Gaussian integral transform and by  expanding  the  integral kernel in terms of Chebyshev polynomials  to allow for a quantification of the theoretical uncertainties.
We perform an analysis of the spectral function and employ it to predict lepton-nucleus scattering.
Our results well describe the $^{16}$O  electron scattering data in the quasi-elastic peak for  momentum transfers $|\mathbf{q}|\gtrapprox500$ MeV and electron energies up to 1.2 GeV, extending therefore the so-called first principles approach to lepton-nucleus cross sections well into the relativistic regime. 
To prove the applicability of this method to neutrino-nucleus cross sections, we implement our $^{16}$O spectral functions in the NuWro Monte Carlo event generator and provide a comparison with recently published T2K neutrino data.

\end{abstract}

\maketitle

\section{Introduction}
Long-baseline neutrino-oscillation programs hold a special place among neutrino experimental endeavours. Driven by their ambitious aims---measuring the CP-violating phase in the electroweak sector and searching for new physics---the next-generation experiments DUNE~\cite{DUNE} and T2HK~\cite{hyperk} will drastically reduce the statistical uncertainty to keep them at the level of a few percent. This implies that systematical uncertainties, until now overshadowed by the statistics, will require more attention. A considerable source of uncertainty comes from the modeling of neutrino-nucleus interactions~\cite{Nustec,whitepaper1,whitepaper2}. 
Given that neutrinos are elusive particles, they can in fact be detected only indirectly via the measurement of the final particles produced in their interaction with target nuclei.
Moreover, oscillation experiments use neutrino fluxes of a wide energy-range, from hundreds of MeV to few GeV, which make them sensitive to a variety of dynamical mechanisms. 
These conditions require an excellent understanding of the underlying  processes to precisely reconstruct the neutrino energy in each observed event.

A theoretical description of electroweak reactions that starts from the forces among nucleons and their interactions with external probes, and is based on a numerical solution of the problem within controlled approximations, is arguably the doorway to a solid understanding of the dynamical mechanism governing lepton-nucleus scattering. This so-called first principles (or ab initio) approach is on the one hand computationally intensive, but on the other hand offers the prospects of quantifying and possibly reducing nuclear physics uncertainties in the computed cross sections.  
For light nuclei, Green Function Monte Carlo (GFMC) has been very successful in delivering predictions of electroweak cross-sections~\cite{Lovato:2017cux,Lovato_2020}.
For nuclei up to mass number 40, the Lorentz integral transform combined with coupled-cluster theory (LIT-CC)  has been recently proven to work well~\cite{Sobczyk:2020qtw,Sobczyk:2021dwm}. Both the above mentioned methods are capable of describing the low energy-momentum part of the lepton-nucleus cross section which depends on the details of the nuclear dynamics in the final state, the so-called final state interaction (FSI). In particular, in the case of the LIT-CC method an extension to higher energies and momenta is complicated by the necessity of using soft nuclear Hamiltonians  which typically have a cutoff of about 500 MeV and therefore cannot be reliably used to describe FSI beyond that point.
However, at higher energies while ground-state correlations remain important, FSIs become negligible and one can assume that only one nucleon in the nucleus interacts with the external probe and is knocked out 
after getting all the momentum transferred from the lepton. In this regime, one can use
 the spectral function (SF) to compute cross sections,  leading to a simplification of the computational task.
 
The SF formalism is a well established approach to describe lepton-nucleus scattering within the impulse approximation (IA). It is based on a factorization ansatz of the ground-state nucleus in terms of one nucleon which participates to the interaction vertex with the external probe, while the remaining $A-1$ nucleus is a spectator.  The SF formalism is  amenable to an extension of the first principle description to the relativistic regime, because one can use relativistic currents in the interaction vertex and even account for higher-energy mechanisms like the pion production~\cite{Rocco:2019gfb}. This is particularly useful in neutrino physics, allowing to address consistently various reactions within the same underlying formalism.

The experimental collaborations perform neutrino energy reconstruction using Monte Carlo (MC) event generators. The analysis requires the knowledge of the semi-inclusive reactions, most importantly the outgoing protons and pions distributions.  Currently, the neutrino community is devoting considerable efforts into improving the quality of implemented nuclear models and into going beyond the simple inclusive cross-sections.
Also in this respect the SF formalism is a convenient tool to model the electroweak processes, because  it allows to address semi-inclusive knockout reactions in a straightforward way. For example,  phenomenological SFs can be constructed using the experimental $(e,e'p)$ data, see, e.g., Ref.~\cite{JeffersonLabHallA:2022cit} for the recent results from the Jefferson Laboratory. 
There are several theoretical  models of phenomenological nature of the spectral function already  available for neutrino studies~\cite{Benhar:1994hw,Nieves:2017lij, Buss:2007ar}. However, with the prospect of quantifying theoretical uncertainties, it is worth investing into the development of spectral functions derived from first principles. A first  calculation based on the self consistent Green's function (SCGF) method was provided in Refs.~\cite{Rocco:2018vbf,Barbieri:2019ual}. Recently,  a new method to construct spectral functions for the many-body system based on Chebyshev polynomials expansion of the integral kernel (ChEK method)~\cite{Roggero:2020qoz,Sobczyk:2021ejs} was proposed. The ChEK method used in conjunction with coupled-cluster theory~\cite{hagen2014} was benchmarked on $^4$He leading to a good agreement with electron-scattering data~\cite{Sobczyk:2022ezo}. In this paper, we present a  computation of the $^{16}$O spectral function with the ChEK method. The main advantage of this approach is that it accounts for the uncertainties of the spectral reconstruction. This feature becomes especially valuable when theoretical uncertainties are propagated to the computed cross sections, allowing for a comparison to experimental data on equal footing. 

The paper is organized in the following way. In Sec.~\ref{sec:lepton-nuc_scattering} we review  the quasi-elastic (QE) process within impulse approximation. In Sec.~\ref{sec:chek} we present the theoretical framework in which we perform the calculation of the spectral functions. In Sec.~\ref{sec:o16} we apply the formalism to $^{16}$O, comparing our results both to electron and neutrino scattering data. The latter is done withing the NuWro Monte Carlo generator~\cite{Juszczak:2005zs,Golan:2012rfa}. Finally, we conclude in Sec.~\ref{sec:conclusion}.

\section{Spectral function formalism in quasi-elastic scattering}
\label{sec:lepton-nuc_scattering}

The differential cross-section for the lepton-nucleus scattering can be expressed as
\begin{equation}
      \frac{d^2\sigma}{d \omega d\cos\theta} =\kappa \frac{|\mathbf{k}|}{|\mathbf{k}'|} L_{\mu\nu} W^{\mu\nu} \,,
\end{equation}
where the energy-momentum transfer is given by $q=(\omega, \mathbf{q})$, the scattering angle $\theta$ and initial and final lepton four-momenta are $k=(E_k,\mathbf{k})$ and $k'=(E_{k'},\mathbf{k'})$, respectively. The interaction vertex depends on the process with
\begin{equation}\kappa_\mathrm{EM}=\left(\frac{\alpha}{q^2}\right)^2\,,~~
\kappa_\mathrm{CC,NC}=\left(\frac{G_F \cos\theta_C}{2\pi}\right)^2 \,,
\end{equation}
for electromagnetic (EM), charge-current (CC) or neutral-current (NC), respectively. The lepton tensor is given by
\begin{equation}
    L_{\mu\nu} = 2a [k_\mu k_\nu' + k_\mu' k_\nu -g_{\mu\nu} (kk')\pm i\eta \epsilon_{\mu\nu\alpha\beta} k^{'\alpha} k^\beta ] \,,
\end{equation}
with $a=1$, $\eta=0$ for electromagnetic and $a=4$, $\eta=1$ for electroweak reactions. The hadronic tensor 
\begin{equation}
    \label{eq:hadronTens}
    W^{\mu\nu} = \sum_f \delta^4(p_0+q-p_f) \langle 0 | \left(J^\mu\right)^\dagger | \Phi_f \rangle \langle \Phi_f | J^\nu |0\rangle \,,
\end{equation}
where  $J_\mu$ is the electroweak current,  and $|0\rangle$ and  $|\Phi_f\rangle$ are the initial  and final nuclear state with respective four momenta $p_0$ and $p_f$,
depends on the reaction mechanism under examination.
For the electroweak current  $J_\mu$, we consider here only one-body operators which in the notation of the second quantization can be written as
\begin{equation}
   J^\mu = \sum_{\alpha,\beta} \langle\beta | j^\mu |\alpha\rangle  a_{\beta}^\dagger a_{\alpha}\,,
   \label{eq:1bcurrent}
\end{equation}
where $\alpha$ and $\beta$  are the quantum numbers of single-particle states created and annihilated by the respective operators $a^{\dagger}$ and $a$.
Within the spectral function formalism, we can use the fully relativistic currents in the matrix element
\begin{equation}
\label{eq:curr}
    \langle p+q | j^\mu | p\rangle = \bar{u}( p+q) \left( V^\mu+A^\mu \right) u( p)\,,
\end{equation}
with Dirac spinors $u$ and the single-nucleon current $j^\mu$ having a vector-axial structure. 
Constructing the most general form of $V^\mu$ and $A^\mu$ using the available four-vectors, we have
\begin{equation}
    \begin{split}
        V^\mu = F_1 \gamma^\mu + \frac{F_2}{2m}i\sigma^{\mu\nu} q_\nu\, , \\
        A^\mu = F_A \gamma^\mu \gamma^5 + \frac{F_P}{m} q^\mu \gamma^5\,,
    \end{split}
\end{equation}
where form factors denoted by $F$ depend on the considered process. For the EM scattering we will use $F_1^{n,p}$, $F_2^{n,p}$ parametrized as in Ref.~\cite{Bradford:2006yz}. The CC vector form-factor is related to the electromagnetic ones as $F_i = F_i^p-F_i^n$. The axial form factors -- present only in the weak interactions -- are related under PCAC (partially conserved axial current):
\begin{equation}
    F_P(Q^2) = \frac{2m^2}{Q^2+m_\pi^2}{F_A(Q^2)}\,,
\end{equation}
with $F_A$ taken as a dipole with $M_A=1030$ MeV axial mass. 

Under the assumption that the struck nucleon does not interact with the final nuclear state, we can factorize the final plane-wave nucleon with momentum $\mathbf{p}'$ and the $A-1$ nuclear state as $|\Phi_f\rangle \to a_{p'}|\Phi_{A-1}\rangle$. By inserting a complete set of intermediate states, $\int d^3\mathbf{p}/(2\pi)^3 |p\rangle \langle p| a_p a^\dagger_p$, the many-body matrix element of the current operator can be approximated by
\begin{equation}
    \begin{split}
        \langle \Phi_f | J^\nu |0\rangle \approx \int \frac{d^3\mathbf{p}}{(2\pi)^3} \langle p'|j^\mu|p \rangle \sum_\alpha \langle \mathbf{p}|\alpha \rangle \langle \Phi_{A-1} |a_\alpha |0\rangle  \,.
    \end{split}
\end{equation}
The hadron tensor factorizes the interaction vertex while the nuclear effects are encapsulated into the spectral function $S(\mathbf{p}, E)$
\footnote{To simplify the notation, from now on we will write $\mathbf{p}$ instead of $|\mathbf{p}|$ when referring to an argument of the SF, $S(\mathbf{p}, E)$, and momentum distribution, $n(\mathbf{p})$.}
, separately for neutrons and protons, as
\begin{align}
W^{\mu\nu}(q)&= \int \frac{d^3\mathbf{p}}{(2\pi)^3} dE \frac{m}{E_p}\frac{m}{E_{p+q}}\nonumber \\
&\big[ S^n(\mathbf{p},E)w_{n}^{\mu\nu}(p,q)+  S^p(\mathbf{p},E)w_{p}^{\mu\nu}(p,q) \big]\nonumber \\
&\times \delta(\omega+E-E_{p+q}-E_f^{kin})\, ,
\label{eq:wmunu-IA}
\end{align}
where $E_f^{kin}$ is the kinetic energy of $A-1$ system and $E_{p+q}$ is the kinetic energy of outgoing nucleon.
Here, the spectral function is defined as 
\begin{equation}
\label{eq:SF}
\begin{split}
    &S(\mathbf{p},E) =  \sum_{\alpha, \beta} \langle \mathbf{p}|\alpha\rangle \langle \mathbf{p}|\beta\rangle^\dagger \\
    &\sumint_{\Phi_{A-1}} \langle 0| a_\beta^\dagger |\Phi_{A-1} \rangle \langle \Phi_{A-1}|  a_\alpha |0\rangle \delta\big(E-(E_0-E_{\Phi})\big)\,.
\end{split}
\end{equation}
and it gives the probability distribution of kicking a nucleon with momentum $|\mathbf{p}|$ out of the ground-state, leaving it with an excitation energy $E$. Finally, the factorized interaction vertex
$w^{\mu\nu}(p,q) =  \langle p+q| j^{\mu} |p\rangle^\dagger \langle p+q | j^{\nu} |p\rangle$ employs the current from  Eq.~\eqref{eq:curr}.


Within the IA, the outgoing nucleon is decoupled from the nuclear final state, hence one is able to factorize the high-energy physics taking place at the interaction vertex from the properties of the nuclear ground state.
However, it is well known that neglecting the FSI at the intermediate momentum transfer of the order of hundreds of MeV leads to some inconsistencies with the data~\cite{Sobczyk:2017mts}. The QE peak is shifted towards higher energy transfers and it exhibits too much strength. These inconsistencies can be partially alleviated if the FSI for the struck nucleon are also included by introducing an optical potential to describe the interaction of the struck nucleon with the rest of the nucleus. Its real part amounts to the potential energy of the nucleon in the nuclear medium, while the imaginary part is responsible to account for the absorption channels of outgoing nucleon.

\begin{figure}[hbt]
    \includegraphics[width=0.5\textwidth]{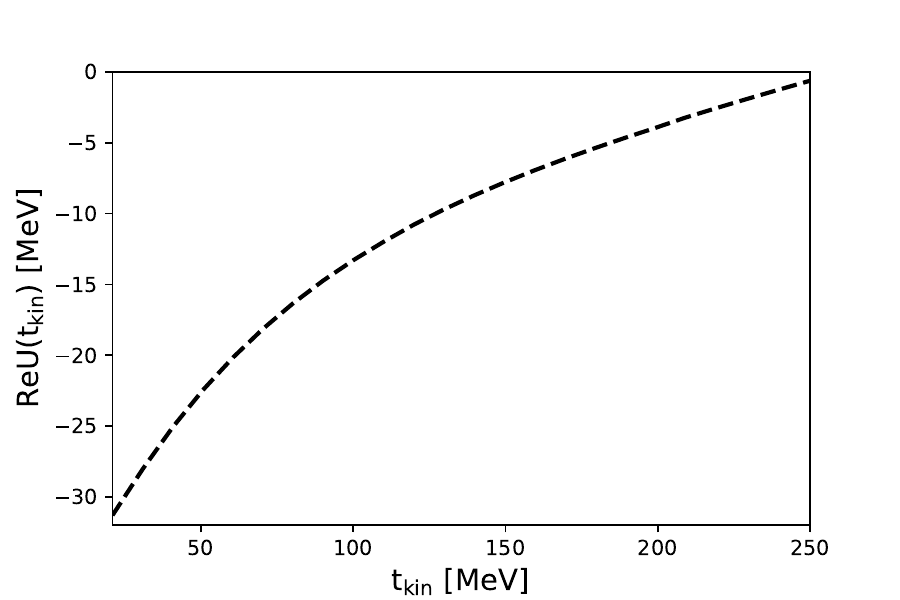}
  \caption{The real part of optical potential for $^{16}$O as a function of the kinetic energy of the outgoing proton from Ref.~\cite{Cooper:1993nx}.}
  \label{fig:optical_pot}
\end{figure}

Phenomenological optical potentials are fitted using elastic nucleon-nucleus scattering data. In the present calculations we will use the real part of the optical potential for $^{16}$O from Ref.~\cite{Cooper:1993nx}. The relativistic potential gives the scalar and vector contributions to the Dirac equation, dependent on the kinetic energy and the radial position. Following the same steps as in Ref.~\cite{Ankowski:2014yfa}, averaging over the density of protons, we arrive at the real part of optical potential $\mathrm{Re}\mathrm{U}(t_\mathrm{kin})$ shown in Fig.~\ref{fig:optical_pot} with $t_\mathrm{kin}$ being the kinetic energy of the outgoing nucleon. The fit of Ref.~\cite{Cooper:1993nx} is reliable only above $t_\mathrm{kin}=25$ MeV (momentum $|\mathbf{p}'|\approx 215$ MeV).
The inclusion of the imaginary part introduced as a folding of the cross section with a Lorentzian function~\cite{Ankowski:2014yfa}, was reported to overestimate the absorption rate leading to non-physical large tails from the Lorentzian distribution, see Ref.~\cite{Benhar:1991af}. Therefore, we will not take the imaginary part into account in our current predictions, since the topic requires further investigations. 

In essence, within our treatment of the FSI, the real part of optical potential enters the hadron tensor changing the energy conservation as
\begin{align}
W^{\mu\nu}_\mathrm{FSI}(q)&= \int \frac{d^3\mathbf{p}}{(2\pi)^3} dE \frac{m}{E_p}\frac{m}{E_{p+q}}\nonumber \\
&\big[ S^n(\mathbf{p},E)w_{n}^{\mu\nu}(p,q)+  S^p(\mathbf{p},E)w_{p}^{\mu\nu}(p,q) \big]\nonumber \\
&\times \delta(\omega+E-E_{p+q}-E_f^{kin}-\mathrm{Re}\mathrm{U})\, .
\label{eq:wmunu-FSI}
\end{align}
Following Ref.~\cite{Ankowski:2014yfa}, we take $\mathrm{Re}\mathrm{U}$ at $t_{kin}=\sqrt{m^2+\mathbf{q}^2}-m^2$.

\section{Spectral functions from the ChEK method}
\label{sec:chek}
Spectral functions as shown in Eq.~\eqref{eq:SF} are defined in terms of the imaginary part of the hole propagator in a many-body system
\begin{equation}
\label{eq:Green}
\begin{split}
    &\mathrm{Im}G_h(\alpha,\beta, E) =\\
    &-\pi \sumint_{\Phi_{A-1}} \langle 0| a_\beta^\dagger |\Phi_{A-1} \rangle \langle \Phi_{A-1}|  a_\alpha |0\rangle \delta\big(E-(E_0-E_{\Phi})\big)\,.
\end{split}
\end{equation}
The spectrum of excited states $\Phi_{A-1}$ contains bound and continuum states, making the direct calculation of $\mathrm{Im}G_h(\alpha,\beta, E)$ challenging. To circumvent this problem we calculate its integral transform
\begin{equation}
\begin{split}
    \mathrm{Im}\tilde G_h(\alpha,\beta, E) &\\
    &=\int d\omega \mathrm{Im}G_h(\alpha,\beta, \omega) K(\omega,E)\\
    &=-\pi \sumint_{\Phi_{A-1}}  \langle 0| a_\beta^\dagger |K\big(E_{\Phi},E-E_0\big) |  a_\alpha |0\rangle \\
    &=-\pi\langle 0| a_\beta^\dagger |K\big( H,E-E_0\big) |  a_\alpha |0\rangle \,,
\end{split}
\label{eq:imGreen_IT}
\end{equation}
with the integral kernel $K(\omega, E)$. If we are interested in $\mathrm{Im}G_h(\alpha,\beta, E)$, the integral transform has to be inverted. The inversion procedure requires in general solving an ill-posed problem. It has been successful for the Lorentz or Laplace kernel when the responses have a relatively simple shape composed of one or two peaks~\cite{Efros:1994iq,Raghavan:2020bze}. A variety of techniques were proposed to achieve it. The spectral function, however, has a more complicated structure, making the inversion practically impossible.
Therefore, here we follow a different strategy outlined in Ref.~\cite{Sobczyk:2022ezo}. We quote here only the most important steps of the derivation, while all the details can be found in Refs.~\cite{Sobczyk:2022ezo,Sobczyk:2021ejs}.

We reconstruct $\mathrm{Im}G_h(\alpha,\beta, E)$ as a histogram.
Our goal is to estimate each bin (centered at $\eta$ having width $2\Delta$), as
\begin{equation}
    \mathrm{Im}G_h(\alpha,\beta; \eta, \Delta) \equiv \int_{\eta-\Delta}^{\eta+\Delta} dE~ \mathrm{Im}G_h(\alpha,\beta, E)\,,
\end{equation}
using the integral transform
\begin{equation}
    \mathrm{Im}\tilde G_h(\alpha,\beta; \eta, \Delta) \equiv \int_{\eta-\Delta}^{\eta+\Delta} dE~ \mathrm{Im}\tilde G_h(\alpha,\beta, E)\,.
\end{equation}
The uncertainty of this reconstruction depends on the properties of the kernel $K$. From our previous studies, we found that the Gaussian kernel has very convenient properties, which we characterize using  parameters $\Sigma$ (accurateness) and $\Lambda$ (resolution)
\begin{equation}
\label{eq:kernel_char}
    \sup_{\omega\in[-1,1]}\sumint_{\omega-\Lambda}^{\omega+\Lambda} K(\omega,E) dE \geq 1-\Sigma\, .
\end{equation}
Using these definitions we arrive at the histogram which is constrained from below and above by the integrated integral transforms, 
\begin{equation}
\label{eq:error_est}
    \begin{split}
 \mathrm{Im}\tilde G_h(\Delta -\Lambda)-\Sigma  \leq\mathrm{Im}G_h(\Delta )  
 \leq \mathrm{Im}\tilde G_h(\Delta +\Lambda)+\Sigma\,,
    \end{split}
\end{equation}
where we suppressed the quantum numbers $\alpha$, $\beta$ and the bin's center $\eta$. 
The integral transform in Eq.~\eqref{eq:imGreen_IT} itself can be calculated in various manners, depending on the employed kernel. For example, the Lorentz integral transform can be conveniently obtained via Lanczos algorithm which gives access to the set of the lowest eigenvalues.
Here, however, we will use a different strategy which can be applied not only to the Lorentzian but also to the  Gaussian kernel, namely by expanding the kernel into Chebyshev polynomials
\begin{equation}
    K(\omega,E) = \sum_{k=0}^N c_k(E) T_k(\omega)\,.
    \label{eq:kernel_expansion}
\end{equation}
We note that Chebyshev polynomials are defined on $[-1,1]$ so we have to scale our problem in such a way that the spectrum of Hamiltonian is confined in $[-1,1]$ range. 
The coefficients $c_k$ of this expansion have analytical form, while the Chebyshev polynomials $T_k$ follow the recursive relations
\begin{equation}
\begin{split}
 &T_0(x) = 1;\, \, \, \, \,T_{-1}(x) = T_1(x) = x;\\
 &T_{n+1}(x) = 2x T_n(x) - T_{n-1}(x)\,.
\label{eq:chebyshev}   
\end{split}    
\end{equation}
We can obtain the moments of this expansion iterating the action of the nuclear Hamiltonian $H$ on the initial state $a_\alpha |0\rangle $
\begin{equation}
\label{eq:cheb_rec}
\begin{split}
    & \langle \tilde \Phi_0| \equiv \langle 0| a_\beta^\dagger| \, ,\ \ \ \ \ |\Phi_0 \rangle \equiv a_\alpha |0\rangle\, ,\\
    &\langle \tilde \Phi_k| \equiv\langle \tilde \Phi_{k-1}|  H \ \ \ \ \ |\Phi_k\rangle =  H |\Phi_{k-1}\rangle\\
    & \mu_0 =  \langle \tilde \Phi_0|\Phi_0 \rangle \, ,\ \ \ \ \ \mu_1 =   \langle \tilde \Phi_0| \Phi_1 \rangle \equiv \langle \tilde \Phi_1| \Phi_0 \rangle \\
    & \mu_{k+1} = 2 \langle \tilde\Phi_0| \Phi_{k+1} \rangle -\mu_{k-1}\equiv 2 \langle \tilde \Phi_{k+1}| \Phi_0 \rangle-\mu_{k-1}\,.
\end{split}    
\end{equation}
Combining Eqs.~\eqref{eq:imGreen_IT} and \eqref{eq:cheb_rec} we arrive at
\begin{equation}
\begin{split}
    \mathrm{Im}\tilde G_h(\alpha,\beta, E) =&-\pi \sum_{k=0}^N c_k(E)  \langle 0| a_\beta^\dagger T_k\big( H\big)   a_\alpha |0\rangle \\
    \equiv&-\pi \sum_{k=0}^N c_k(E) \mu_k \,.
\end{split}
\label{eq:imGreen_kernel}
\end{equation}
We truncate the expansion at the level on $N$ moments, introducing a controllable  error $\gamma$,
\begin{equation}
    \gamma = \sum_{k=N+1}^\infty c_k(E) T_k(\omega)\,,
    \label{eq:trunc_error}
\end{equation}
It has to be included into an overall uncertainty budget, leading to the final prediction
\begin{equation}
\label{eq:error_est}
    \begin{split}
 \mathrm{Im}\tilde G_h(\Delta -&\Lambda)-\Sigma-2\gamma(\Delta -\Lambda)  \\
 &\leq\mathrm{Im}G_h(\Delta )  \\
 \leq \mathrm{Im}\tilde G_h&(\Delta +\Lambda)+\Sigma+2\gamma(\Delta +\Lambda)\,.
    \end{split}
\end{equation}
By setting $\Delta$ (the histogram's width), $\Lambda$ (width of the kernel) and $N$ (number of Chebyshev moments) we can estimate the lower and upper bound on $\mathrm{Im}G_h(\Delta)$ according to Eq.~\eqref{eq:error_est}. We note that $\Sigma$ and $\gamma$ have a known analytical form, and the uncertainty is mainly driven by $|\mathrm{Im}\tilde G_h(\Delta +\Lambda)-\mathrm{Im}\tilde G_h(\Delta -\Lambda)|$.

\subsection*{Coupled-cluster theory}
We calculate the Chebyshev moments $\mu_k$ within the spherical coupled-cluster framework~\cite{hagen2014}. This formalism starts from a reference state $|\Phi\rangle$, in our case a Hartree-Fock solution, on top of which we include the nuclear correlations using an exponential ansatz
\begin{equation}
\begin{split}
\label{eq:excitationOperator}
&|0\rangle = e^T |\Phi\rangle\,, \\
&T=\sum_{i,a} t^a_i a^\dagger_a a_i + \frac{1}{4}\sum_{ijab}t^{ab}_{ij}a_a^\dagger a_b^\dagger a_i a_j +...\,.
\end{split}
\end{equation}
In our present calculation we retain the first two terms of this expansion, i.e. we work in the singles and doubles (CCSD) approximation. 
The $t$ amplitudes appearing in the correlation operator $T$ can be determined solving a set of coupled nonlinear equations. 
For the calculation of the Green's function we need to construct a set of initial $A-1$ states acting with the similarity transformed  operators $\overline{a}_\alpha$ and $\overline{a}^\dagger_\alpha$ on the left and right ground-state
\begin{equation}
\begin{split}
    &|\Phi_0\rangle = \overline{a}_\alpha|\Phi\rangle \equiv e^{-T} a_\alpha e^T |\Phi\rangle \,,\\
    &\langle\tilde\Phi_0| = \langle \Phi| \overline{a}^\dagger_\alpha \equiv \langle \Phi|(1+\Lambda) e^{-T} a^\dagger_\alpha e^T\, ,
\end{split}
\end{equation}
where $\Lambda$ is the de-excitation operator which has to be included since the coupled-cluster is a non-hermitian theory yielding different left and right eigenstates.
The calculation of Chebyshev moments requires an iterative action of the similarity transformed Hamiltonian, following the recursive relation from Eq.~\eqref{eq:cheb_rec}. 
%

%
%
%
%
%
%

\section{Results}
\label{sec:o16}

In all the results presented in this paper, we employ the NNLO$_\mathrm{sat}$ nuclear Hamiltonian~\cite{Ekstrom:2015rta} containing both nucleon-nucleon (NN) and three-nucleon (3N) forces derived in chiral effective field theory at next-to-next-to leading order~\cite{Epelbaum:2008ga}. The low-energy constants in this Hamiltonian are fitted both to NN scattering data, as well as to properties of light nuclei and selected medium-mass nuclei. We recall that the current operators implemented in this work are not derived in  chiral effective field theory, but we rather use the relativistic forms described in Section~\ref{sec:lepton-nuc_scattering}.
In the coupled-cluster calculations, 3N interactions are approximated at the normal-ordered two-body level~\cite{Hagen:2007ew,Roth:2011vt}, and an additional cut on three-nucleon configurations $E_{3max} \leq 16$ is imposed. We performed calculations for the model space of 15 oscillator shells and values of underlying harmonic oscillator frequencies $\hbar \Omega=12-20$ MeV.

\begin{figure}[hbt]
    \includegraphics[width=0.45\textwidth]{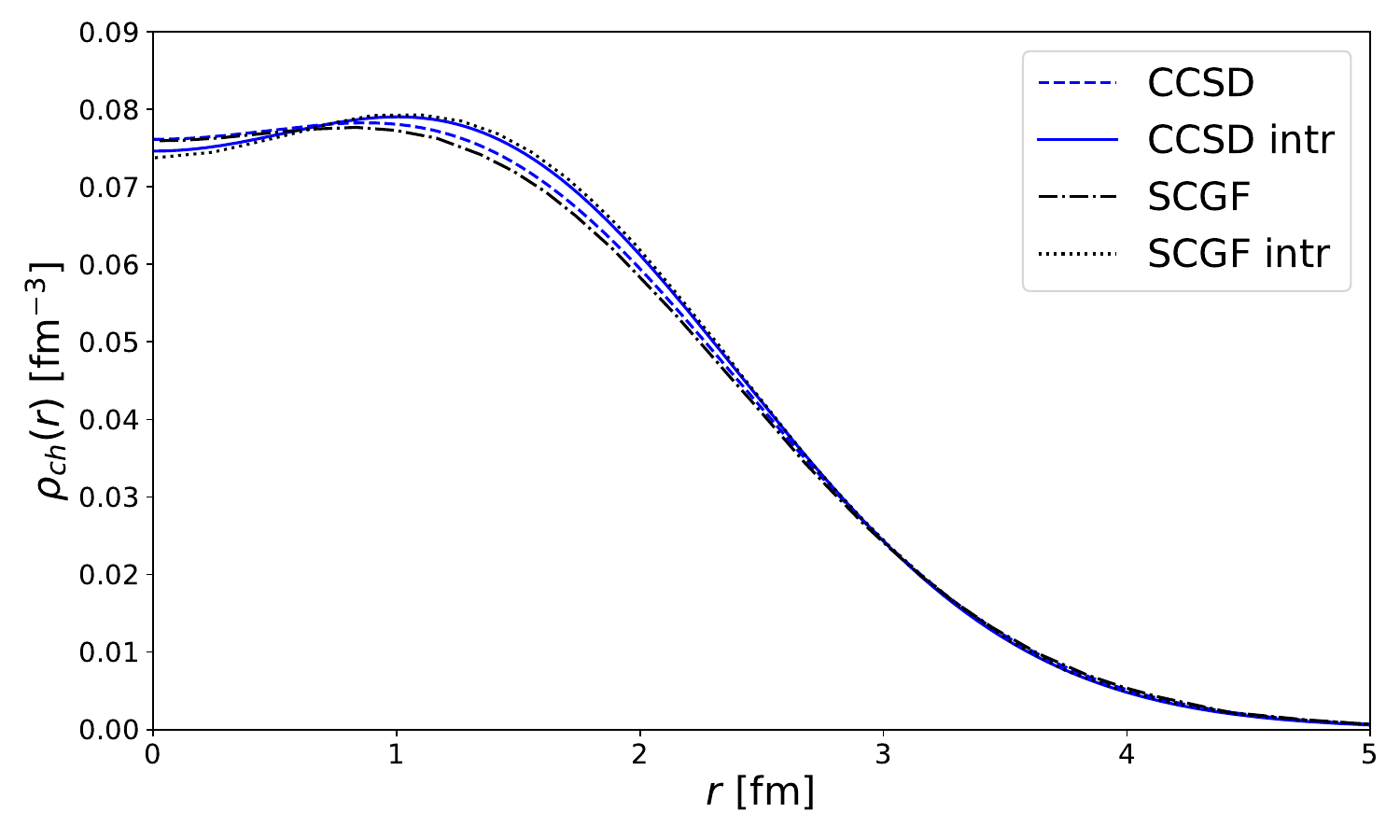}
    \includegraphics[width=0.45\textwidth]{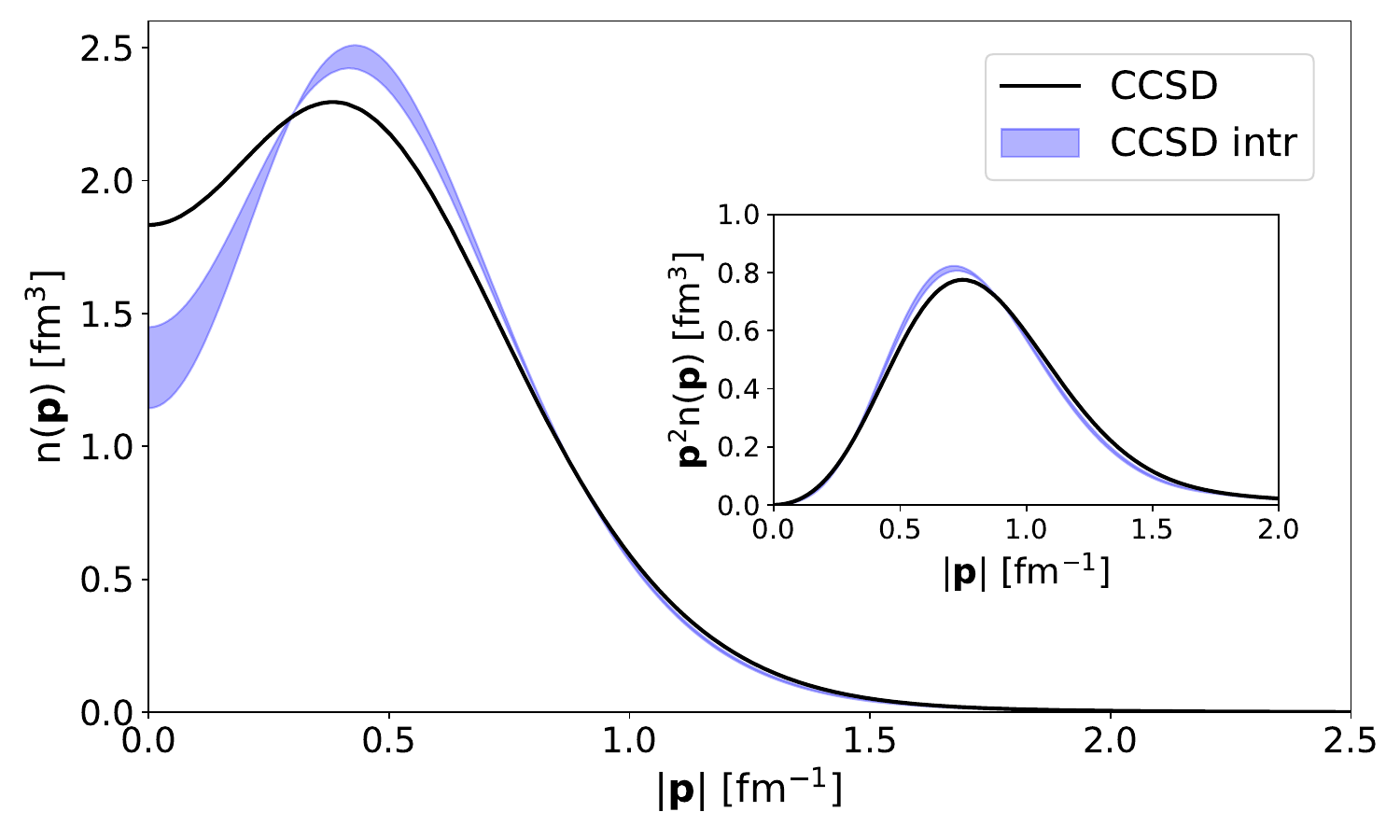}
  \caption{Charge distribution (upper panel) comparison between CCSD and SCGF calculations using the same interaction NNLO$_\mathrm{sat}$. Momentum distribution (lower panel) using CCSD. See text for details.}
  \label{fig:momentum_distr}
\end{figure}

To benchmark our calculation we first  look at the charge distribution and compare it with previous results from the SCGF~\cite{Rocco:2018vbf} method. In the upper panel of Fig.~\ref{fig:momentum_distr} we present the direct result of the computation which includes spurious center of mass (CoM) contaminations, denoted with CCSD. We also show the intrinsic charge distribution, for which the CoM contributions were subtracted (see Ref.~\cite{Sobczyk:2022ezo} for details),  denoted with ``CCSD intr''. They are both in a very good agreement with the SCGF predictions, for which a different numerical procedure is used to remove the CoM contributions. We also note that the ``CCSD'' and ``CCSD intr'' distributions are similar, which confirms the well known fact that spurious CoM effects decrease with the nuclear mass (for a comparison see Fig.~2 in Ref.~\cite{Sobczyk:2022ezo} for $^4$He, where the effect was larger). 

Next, we look at the momentum distribution $n(\mathbf{p})=\int d E\, S(\mathbf{p}, E)$ to further assess the role of the CoM contamination
We follow the same procedure as explained in Ref.~\cite{Sobczyk:2022ezo} to calculate the intrinsic $n(\mathbf{p})$. In Fig.~\ref{fig:momentum_distr} we show both $n(\mathbf{p})$ and $\mathbf{p}^2 n(\mathbf{p})$ in the inset. The CoM removal affects mostly low momenta (shown as a blue band in Fig.~\ref{fig:momentum_distr}). The uncertainty comes from varying the width of the CoM Gaussian used as an ansatz, $\hbar\tilde\Omega=16-24$ MeV. The CoM effect will be negligible when we consider the momentum-weighted $\mathbf{p}^2 n(\mathbf{p})$, as in the case of cross-section calculation. We will therefore safely neglect this effect from now on. 

\begin{figure}[hbt]
    \includegraphics[width=0.49\textwidth]{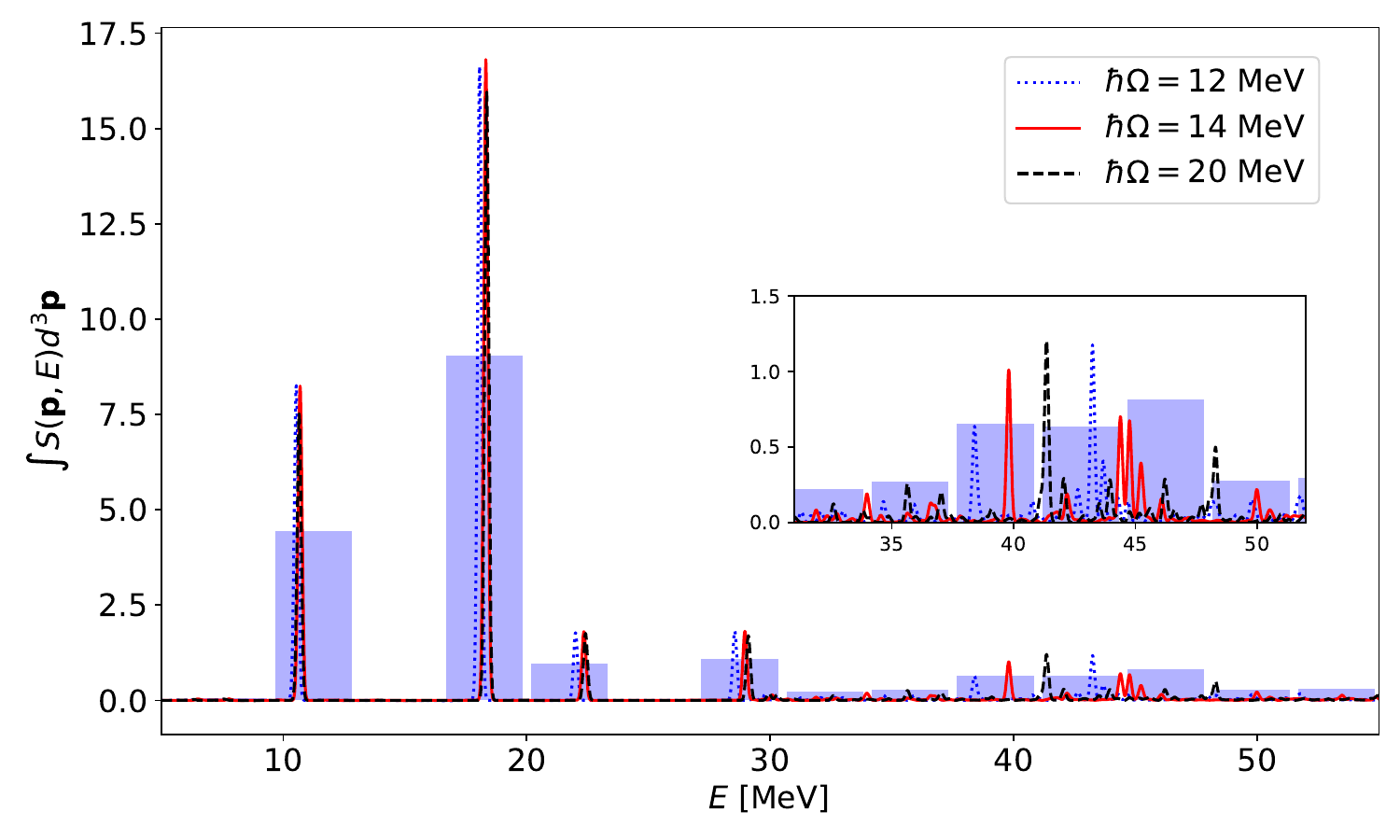}
  \caption{The integrated energy distribution $S(E)$ of the proton spectral function for various values of $\hbar \Omega=12-20$ MeV. The chosen binning is also shown (arbitrary normalization).  }
  \label{fig:o16_hw}
\end{figure}

In order to investigate the dependence of the spectral function on the basis, we looked separately at integrated distributions: $n(\mathbf{p}) = \int dE\ S(\mathbf{p},E)$  and $S(E)=\int d^3\mathbf{p}\ S(\mathbf{p},E)$.
Momentum distribution $n(\mathbf{p})$ is practically independent on the choice of $\hbar\Omega$. 
The differences between various $S(E)$ can be appreciated in Fig.~\ref{fig:o16_hw}. The dominating peaks (below $30$ MeV) have almost the same strength and are shifted by less than $1$ MeV, while the spectrum above 30 MeV is quite different, as can be seen in the inset of Fig.~\ref{fig:o16_hw}.
A direct comparison of the full 2D distribution of the spectral function for various $\hbar \Omega$ reveals some strength redistribution. There are also some small regions which give a negative contribution. This non-physical behaviour is most likely due to the fact that the coupled-cluster theory is non-hermitian. The appearance of small admixtures of non-physical states has been already observed~\cite{Gu:2023aoc,Hagen:2010gd}. 
We have numerically checked that the negative contribution is smallest for $\hbar \Omega=14$ MeV and in this case it stays at the per-mil level. For other values of oscillator frequencies it reaches at most $4\%$ for $\hbar \Omega=20$ MeV. Therefore, we have decided to perform all the further calculations with this optimal value of $\hbar \Omega=14$ MeV which alleviates the non-physical behaviour. 

\begin{figure}[hbt]
    \includegraphics[width=0.49\textwidth]{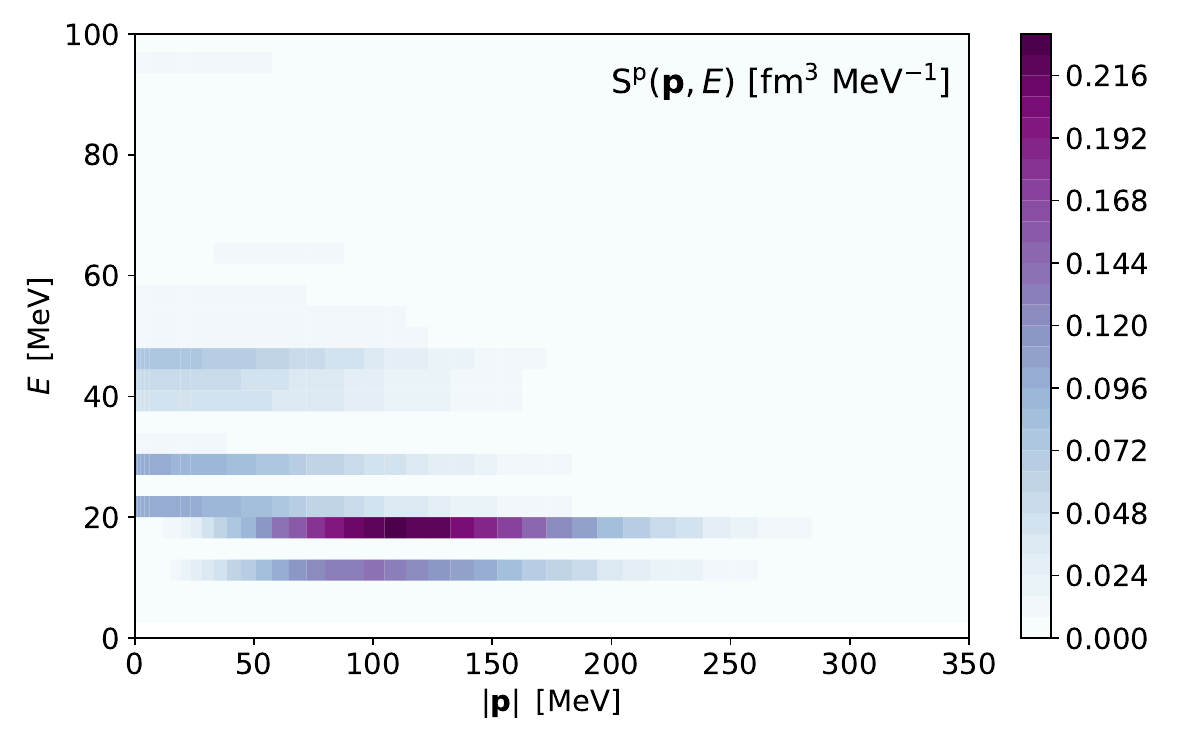}
    \includegraphics[width=0.49\textwidth]{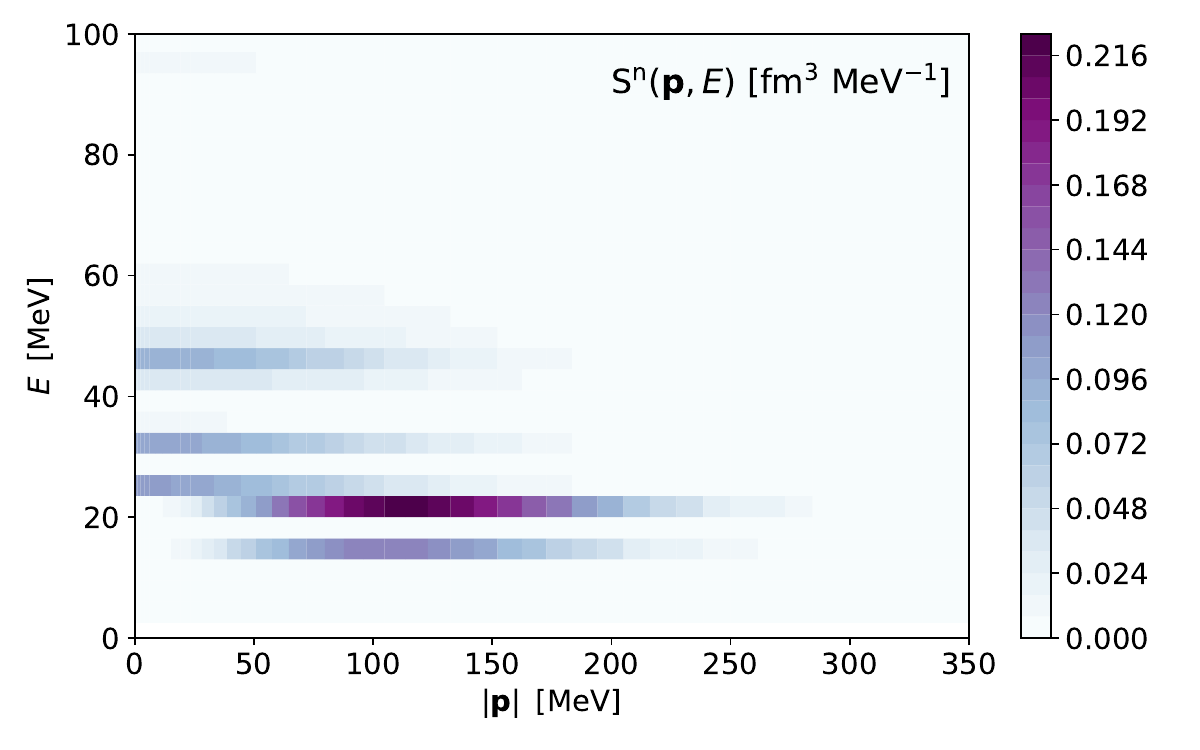}
  \caption{$^{16}$O spectral functions for protons (upper panel) and neutrons (lower panel). See text for details of spectral reconstruction. The theoretical uncertainty, not shown in this figure, is discussed in Subsec.~\ref{sec:uncertainty}.}
  \label{fig:sf_o16}
\end{figure}

In Fig.~\ref{fig:sf_o16}, we show the final SF histograms separately for protons and neutrons using $2\Delta=3.5$ MeV binning. In this case the Hamiltonian spectrum is limited by energy $E=250$ MeV.\footnote{This value is needed to scale the spectrum to $[-1,1]$ range where the Chebyshev polynomials are defined.} We set $\Lambda=0.66$ MeV, and the number of Chebyshev moments $N=4000$ to keep the truncation error $\gamma$ negligibly small. One can observe two clearly dominating peaks at $E\approx13(10)$ and $21(18)$ MeV for neutrons (protons), which correspond to $1p_{1/2}$ and $1p_{3/2}$ states, and some strength distributed at higher energies. We note that for this estimation we use $\mathrm{Im}G_h(\Delta)\approx \mathrm{Im}\tilde G_h(\Delta)$ which lies between the lower and upper bounds, as shown in Eq.~\eqref{eq:error_est}. 

\subsection{Uncertainty estimation}
\label{sec:uncertainty}

\begin{figure*}[hbt]
    \includegraphics[width=0.325\textwidth]{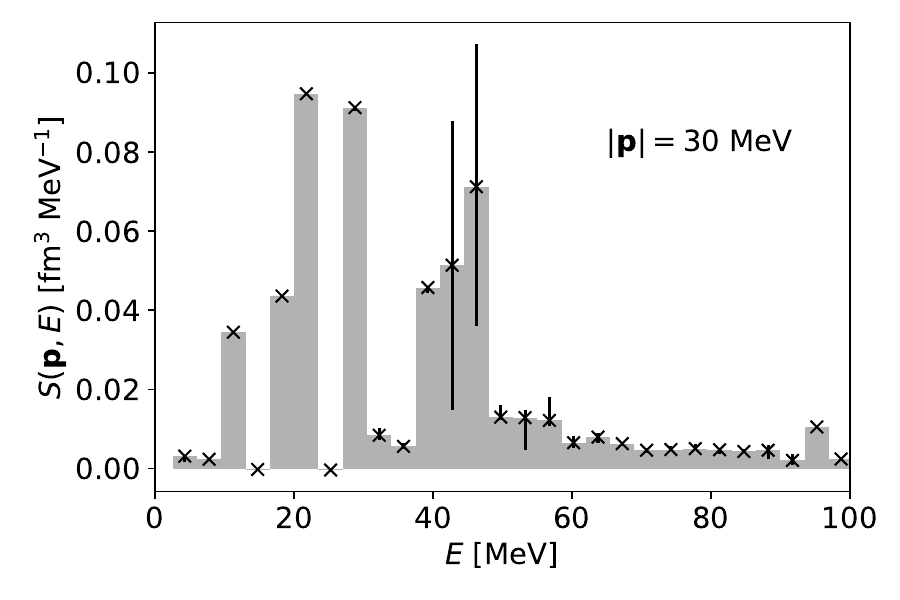}
    \includegraphics[width=0.325\textwidth]{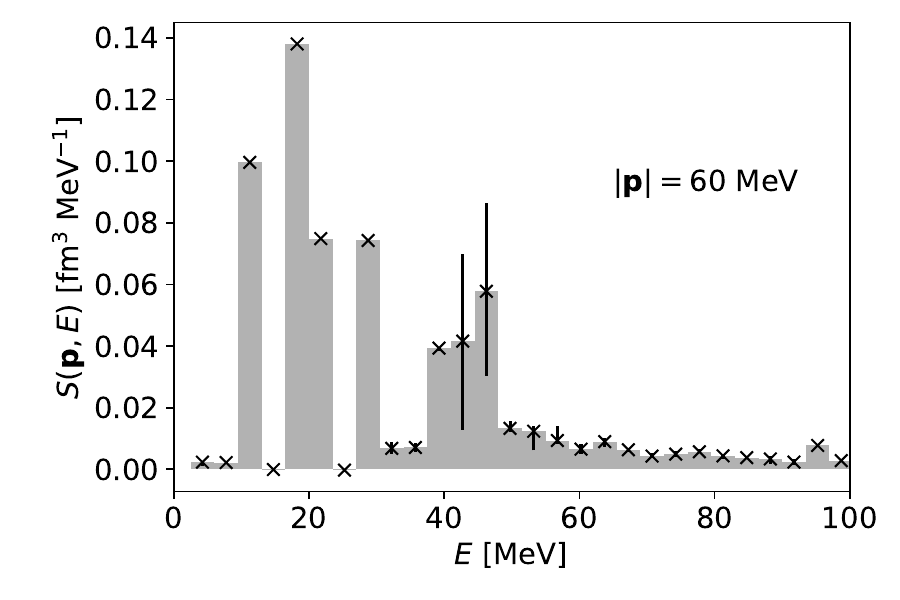}
        \includegraphics[width=0.325\textwidth]{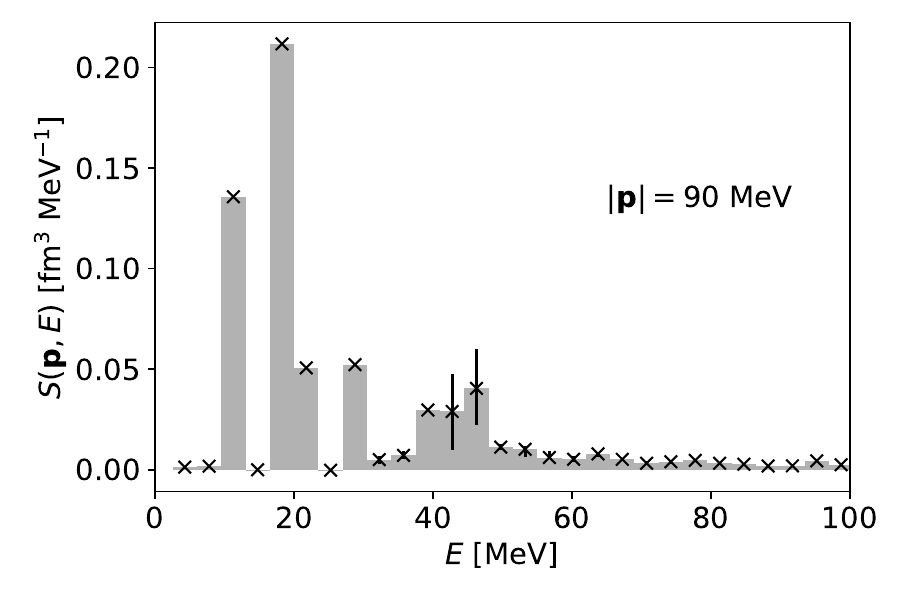}
  \caption{Proton spectral function of $^{16}$O for three values of momentum $|\mathbf{p}|$. The uncertainty bars come from Eq.~\eqref{eq:error_est}.}
  \label{fig:sf_o16_errors}
\end{figure*}

The estimated uncertainties coming from the ChEK procedure (see Eq.~\eqref{eq:error_est}) affect mostly the energy range $E>40$ MeV. 
This can be understood when various sources of uncertainty are analyzed in Eq.~\eqref{eq:error_est}. With our choice of parameters we keep $\Sigma$ and $\gamma$ small, and the uncertainty is driven by $| \mathrm{Im} \tilde G_h(\Delta+\Lambda)-\mathrm{Im} \tilde G_h(\Delta-\Lambda)|$. The lower part of the spectrum (below 30 MeV) is composed of well separated peaks, and therefore with our choice of the histogram binning the uncertainties are negligible. The total strength of the SF is dominated by this region. Therefore, the uncertainties have an overall small impact on the cross section.
The uncertainties estimated for $\mathrm{Im}\tilde G(\alpha,\beta, E)$ lead to $|\mathbf{p}|$-dependent errors in the final SF, according to  Eq.~\eqref{eq:SF}.
In Fig.~\ref{fig:sf_o16_errors} we show the proton spectral function at three values of momenta $|\mathbf{p}|$. The spectrum below $E=40$ MeV is not affected by the uncertainties while for higher energies the errors are larger. They will not, however, influence much the cross-section results, since the hadron tensor is weighted by $\mathbf{p}^2 d|\mathbf{p}|$. In fact, the weighted momentum distribution $\mathbf{p}^2n(\mathbf{p})$ peaks at $|\mathbf{p}|\approx 150$ MeV (see Fig.~\ref{fig:momentum_distr}).

\subsection{Applications to electron-nucleus scattering}
Electron scattering experiments serve as an excellent test to check the reliability of the nuclear models and of the assumed approximations. Unfortunately, there are only scarce data available for $^{16}$O,  corresponding to  momentum transfers in the range of $320-650$ MeV. 
\begin{figure*}[hbt]
    \includegraphics[width=0.45\textwidth]{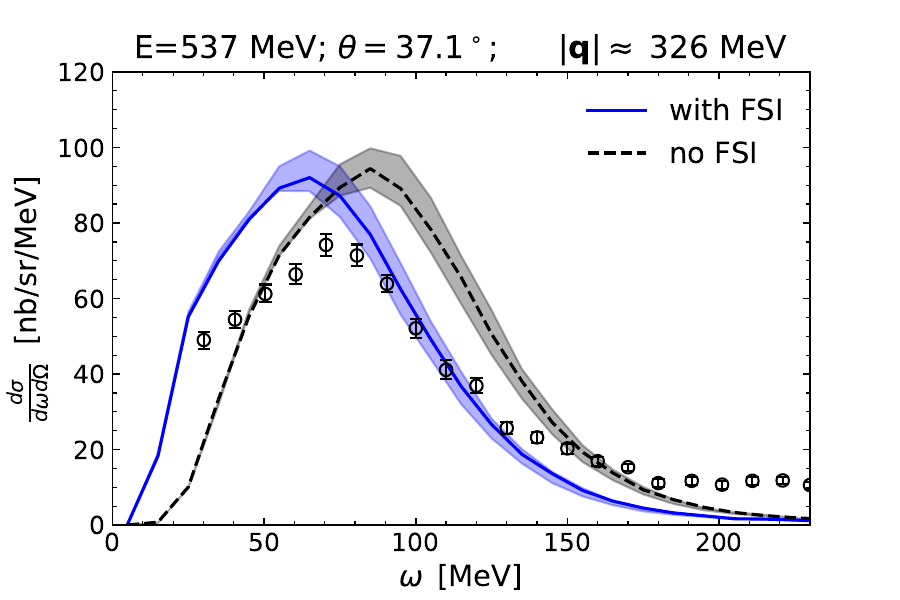}
    \includegraphics[width=0.45\textwidth]{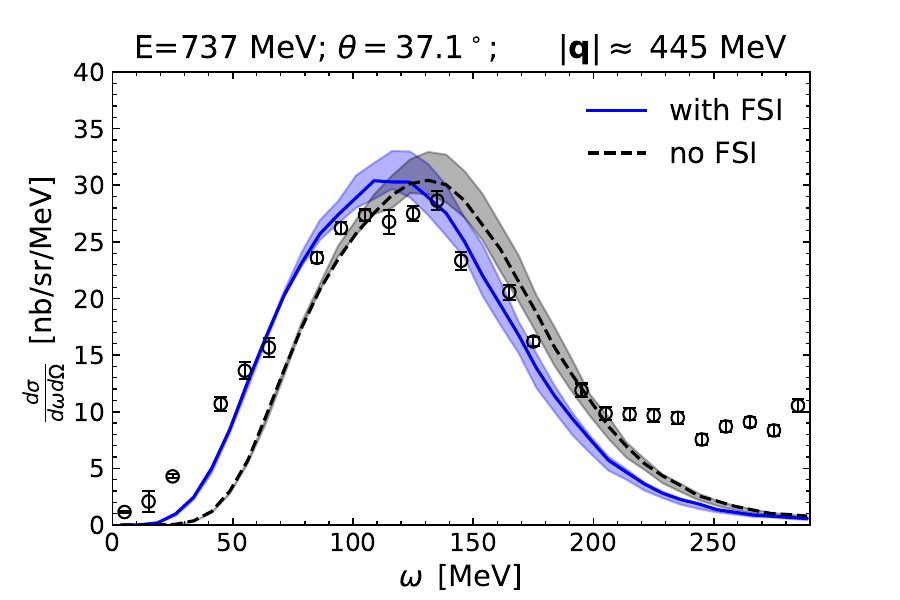}
    \includegraphics[width=0.45\textwidth]{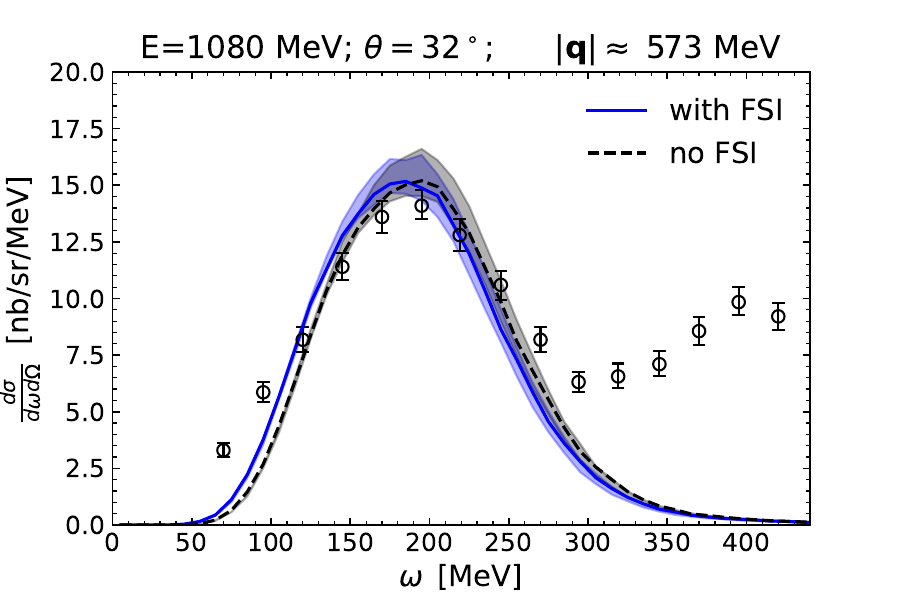}
    \includegraphics[width=0.45\textwidth]{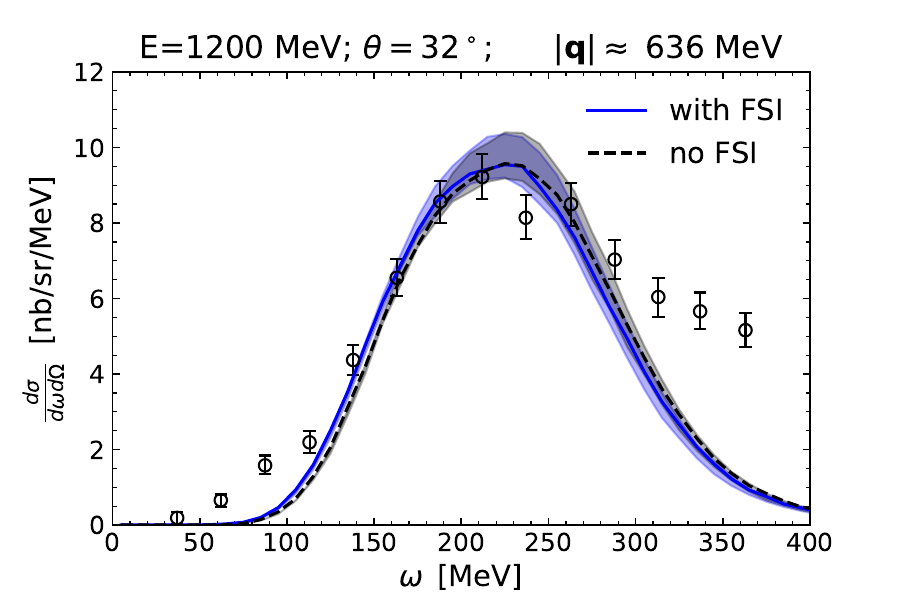}
\caption{Electron scattering differential cross-section on $^{16}$O for different kinematics which correspond to the momentum transfer $|\mathbf{q}|\approx 320-650$ MeV. We show results obtained with CCSD without including the optical potential (dashed line) and after its inclusion (solid line). The uncertainty bands come from the uncertainty of the ChEK method. Experimental data was taken from Ref.~\cite{OConnell:1987ag, Anghinolfi:1996vm}.}
  \label{fig:o16_electron_scattering}
\end{figure*}
In Fig.~\ref{fig:o16_electron_scattering}, we compare our results with the available experimental data. Clearly, our calculations compare better in the kinematical regime of higher momentum transfer, as expected from an IA assumption. When accounting for the FSI via the optical potential as described in Sec.~\ref{sec:lepton-nuc_scattering},  the  position of the QE peak is shifted to lower energy transfers so that the agreement with the experimental data improves. As anticipated, the effect of FSI is stronger in the lower momentum-energy regimes presented in the first row of Fig.~\ref{fig:o16_electron_scattering}. 

Within our approach, we propagate the theoretical uncertainty from the SF to the cross-section results. In practice, we construct two SFs taking both the lowest and the highest values for each histogram bin as presented in Fig.~\ref{fig:sf_o16_errors}, and with those two SFs we construct a lower and an upper  cross section, respectively, which lead to the bands in Fig.~\ref{fig:o16_electron_scattering}. The obtained uncertainty is of the order of a few percents and reaches about the $10\%$ mark at the QE peak.
Although the response in the QE peak for $|\mathbf{q}|>500$ MeV seems to  well describe the data, our predictions might actually be too high. In fact, other mechanisms not included in the calculation, such as meson exchange currents (MEC) and pion production, typically contribute by mostly enhancing the high-energy slope of the QE peak.
However, we note that the imaginary part of the optical potential, not included in our current calculations, could quench the response, giving therefore some room for the above mentioned enhancing contributions. Therefore, we expect here some cancellations of the omitted contributions, whose investigation is left to future work.

\begin{figure}[hbt]
    \includegraphics[width=0.45\textwidth]{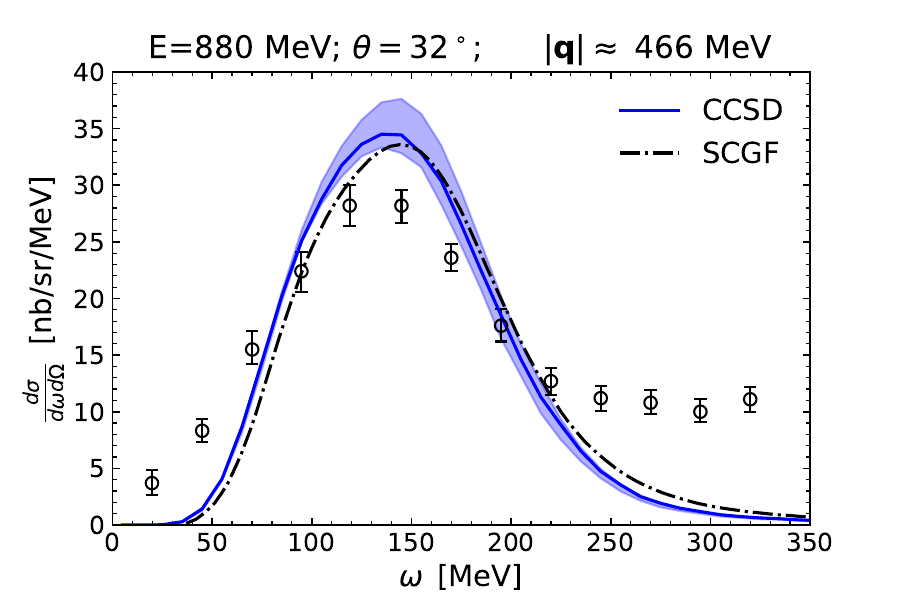}

  \caption{Electron scattering cross section for $^{16}$O calculated within CCSD (continuous blue line) and within SCGF~\cite{Rocco:2018vbf} (black dashed-dotted line)  in comparison to experimental data from Ref.~\cite{Anghinolfi:1996vm}.}
  \label{fig:o16_scgf_comp}
\end{figure}

Finally, in Fig.~\ref{fig:o16_scgf_comp} we  present a comparison of our calculations with the previous results of Ref.~\cite{Rocco:2018vbf} obtained from the SCGF method. We show only one kinematics, since the  comparison is similar for other setups. We observe that the two curves are very similar, indicating a nice agreement. Looking at the details, the QE peak in CCSD is slightly shifted towards the left with respect to the SCGF calculation. We expect that the source of this tiny deviation lies in the differences between the two many-body methods. 
As pointed out before, we obtain also very similar charge distributions. We have checked that the  momentum distributions are also in very good agreement between the two methods, indicating that the many-body description of the ground states is very similar. However, the slight differences observed in the cross section may indicate a stronger sensitivity to the details of the many-body method for dynamical observables. We nevertheless consider this benchmark very successful.

\subsection{Applications to neutrino-nucleus scattering}
We now redirect our attention to the application of our calculations to neutrino-nucleus scattering.
The $^{16}$O spectral function is of particular interest for T2K and future T2HK experiments.
A direct comparison of the QE peak with the data for the neutrino-nucleus scattering is currently not possible due to the experimental constraints. The neutrino flux has a broad energy spectrum so that many mechanisms contribute and cannot be well separated. Moreover, the current experimental uncertainties are large and dominated by the statistics. Nevertheless, recently the T2K collaboration published inclusive cross-section on $^{16}$O for CC$0\pi$ events (no pions detected in the final state) for various angles of the outgoing muon~\cite{T2K:2020jav}. This observable should have a large contribution coming from the single-nucleon knockout mechanism, which is well described by our SF. To make a full comparison with the data, we need to $(i)$ account for all other possible mechanisms, beyond the one-nucleon knockout and $(ii)$ get the distribution of the produced hadrons going beyond the inclusive cross-section. We achieve this by implementing our SF in the NuWro MC event generator~\cite{Juszczak:2005zs,Golan:2012rfa}.
Typically, the MC generators describe the neutrino-nucleus scattering in a two-step process. In the first step, the scattering takes place on a single nucleon (or a pair of nucleons in case of meson-exchange currents) in the primary vertex. This is where we include our SF model. In the next step, the produced particles (predominantly nucleons and pions) are cascaded through the nucleus, where they can re-scatter, be absorbed, or produce other particles. Various approaches to model the inter-nuclear cascade were recently compared in Ref.~\cite{Dytman:2021ohr}. From this perspective including the imaginary part of the optical potential -- omitted in our calculation --  might lead to the double counting of some effects already accounted for in the cascade.

\begin{figure*}[hbt]
    \includegraphics[width=0.45\textwidth]{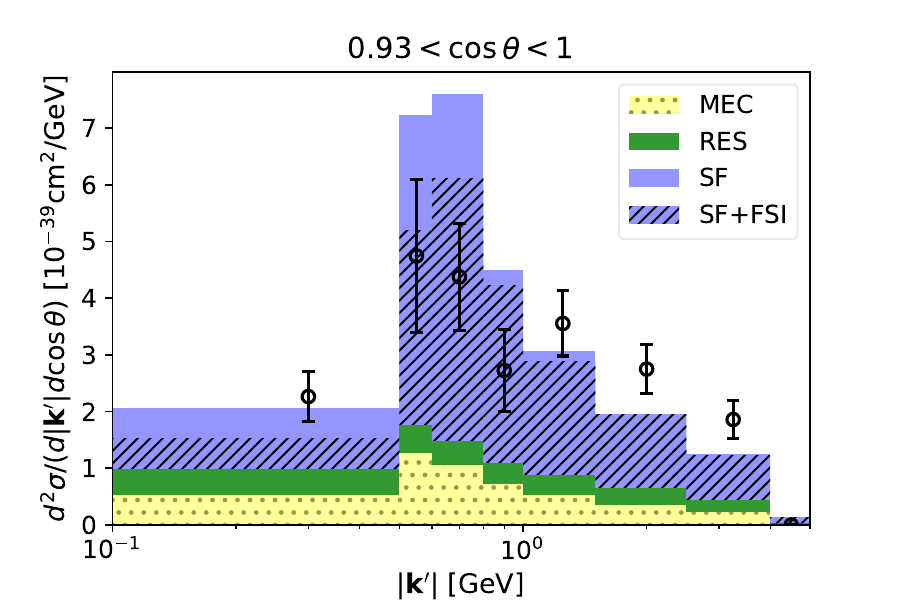}
    \includegraphics[width=0.45\textwidth]{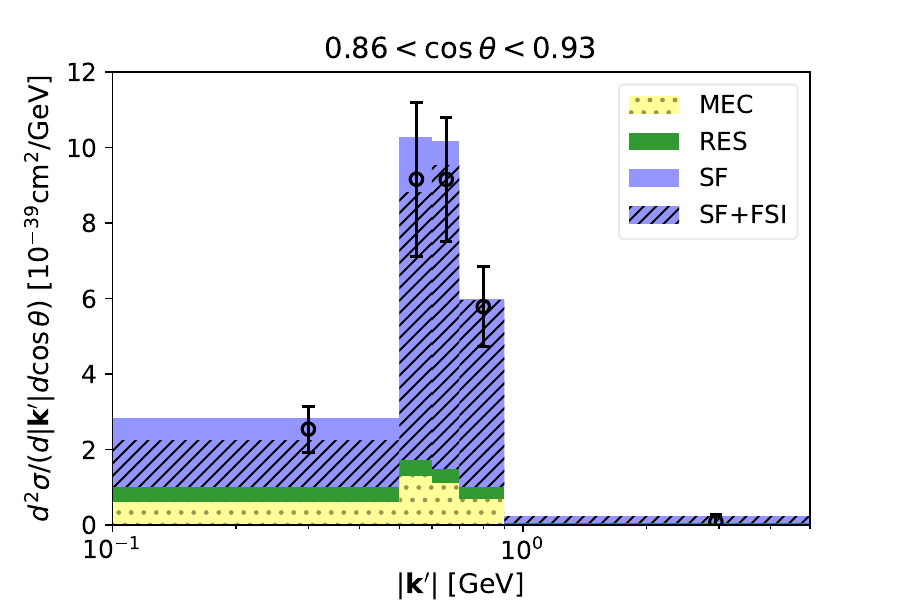}
    \includegraphics[width=0.45\textwidth]{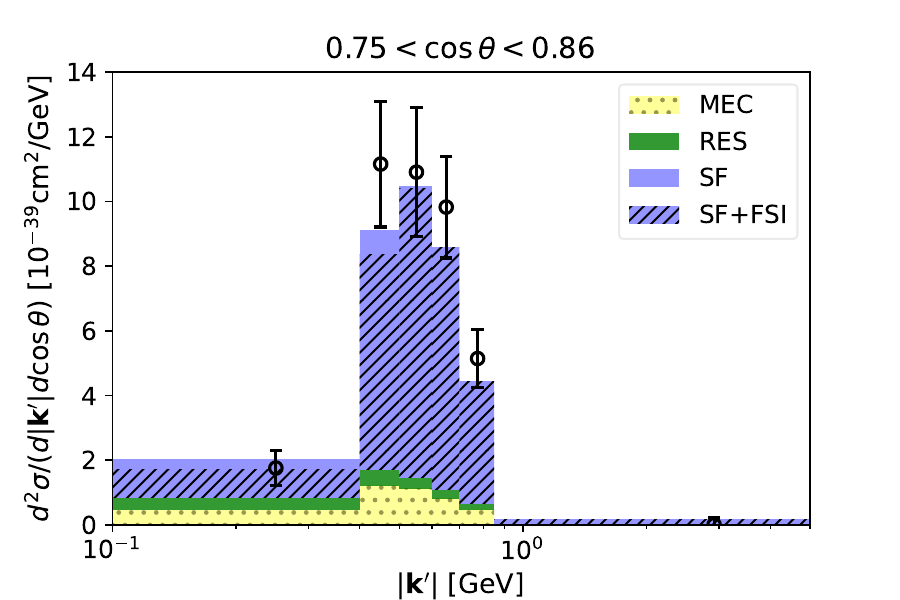}
    \includegraphics[width=0.45\textwidth]{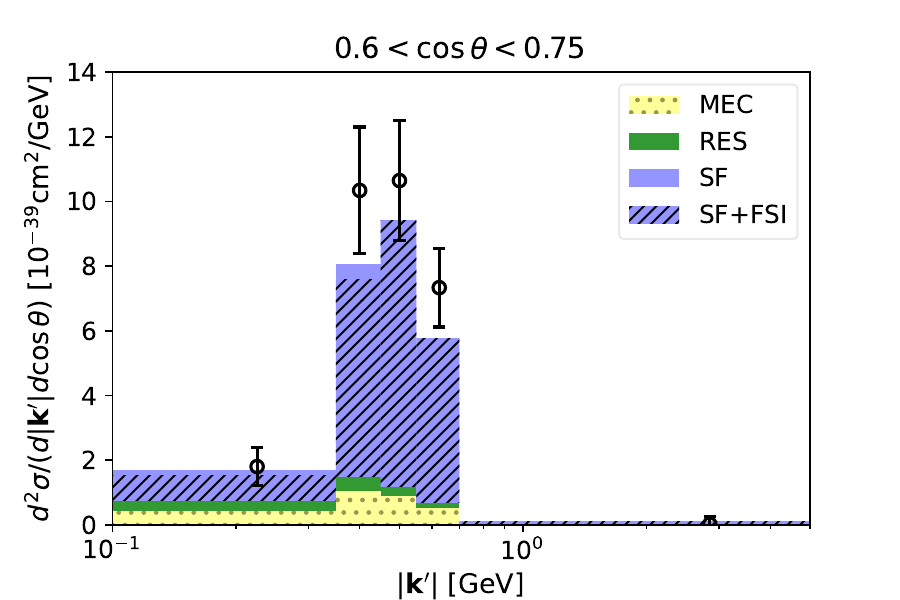}
    \includegraphics[width=0.45\textwidth]{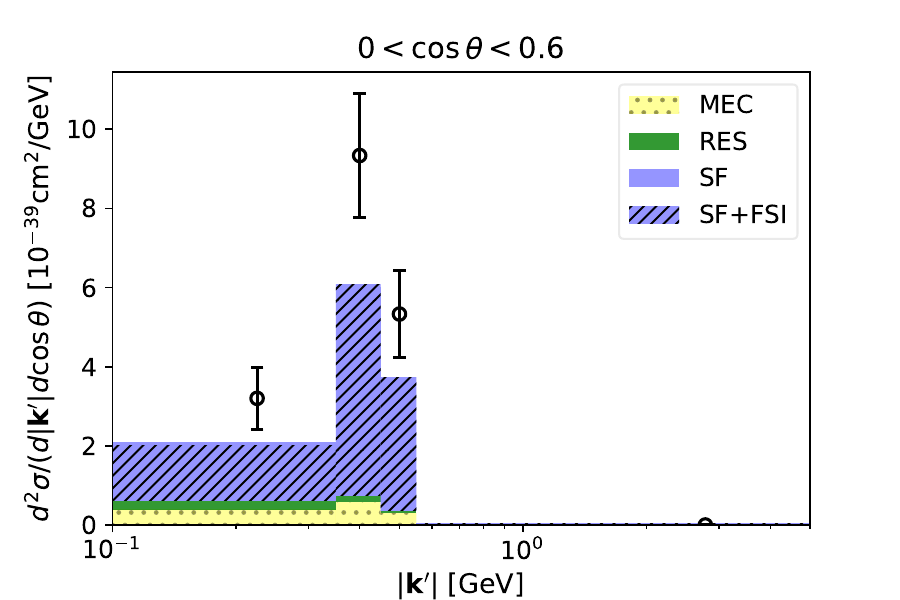}

  \caption{The double differential cross section $\nu_\mu + ^{16}\mathrm{O} \to \mu^- + X$ of CC0$\pi$ events measured by the T2K experiment~\cite{T2K:2020jav}. The momentum distribution $|\mathbf{k}^\prime|$ of outgoing $\mu^-$ was measured for five ranges of the scattering angle $\cos\theta$. The results ``SF'' were obtained using our spectral function, while ``SF+FSI'' include also the real part of the optical potential. In both cases we do not include theoretical uncertainties. Other mechanisms, predominantly MEC and resonance production (RES) give smaller contribution according to the model used in the simulation~\cite{T2K:2020jav}. }
  \label{fig:t2k}
\end{figure*}

\begin{figure*}[hbt]
    \includegraphics[width=0.45\textwidth]{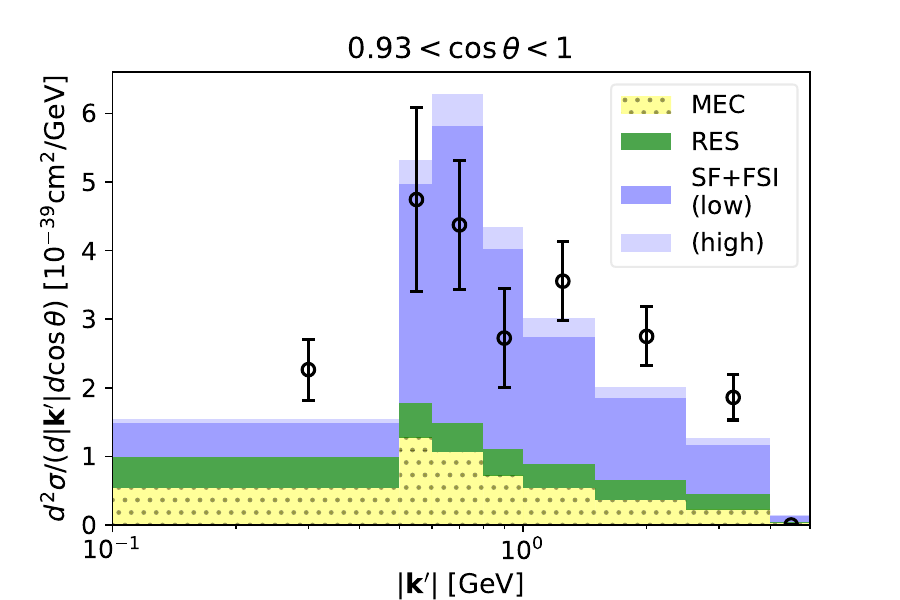}
    \includegraphics[width=0.45\textwidth]{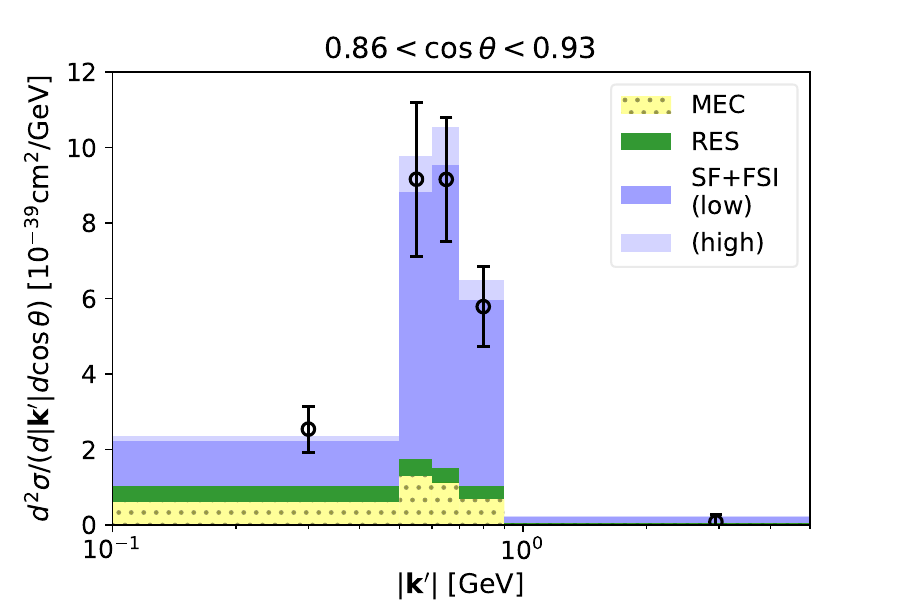}
    \includegraphics[width=0.45\textwidth]{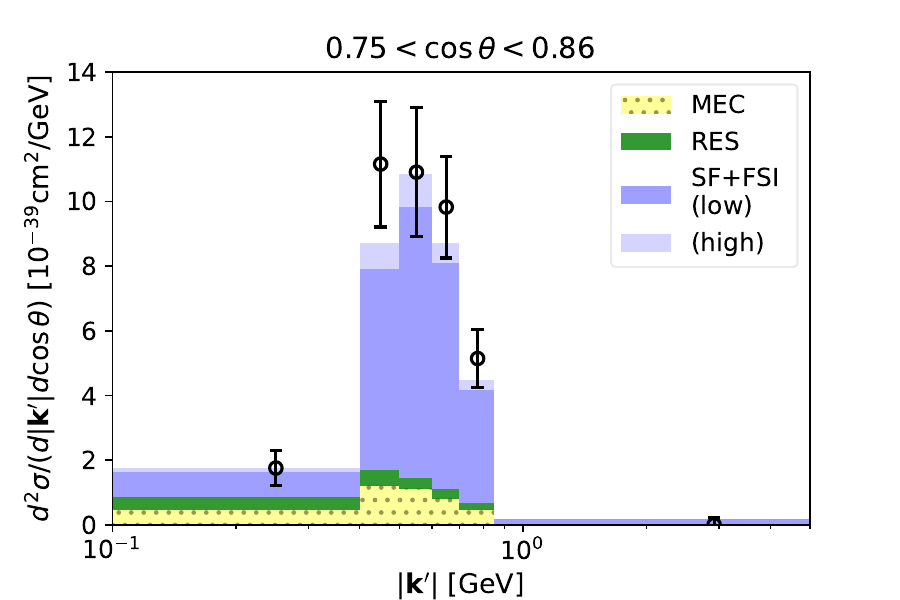}
    \includegraphics[width=0.45\textwidth]{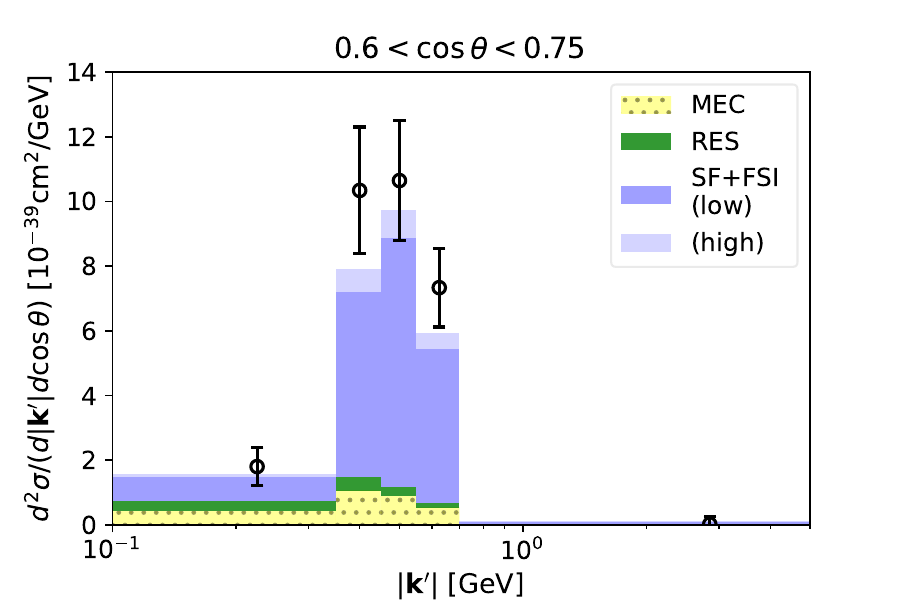}
    \includegraphics[width=0.45\textwidth]{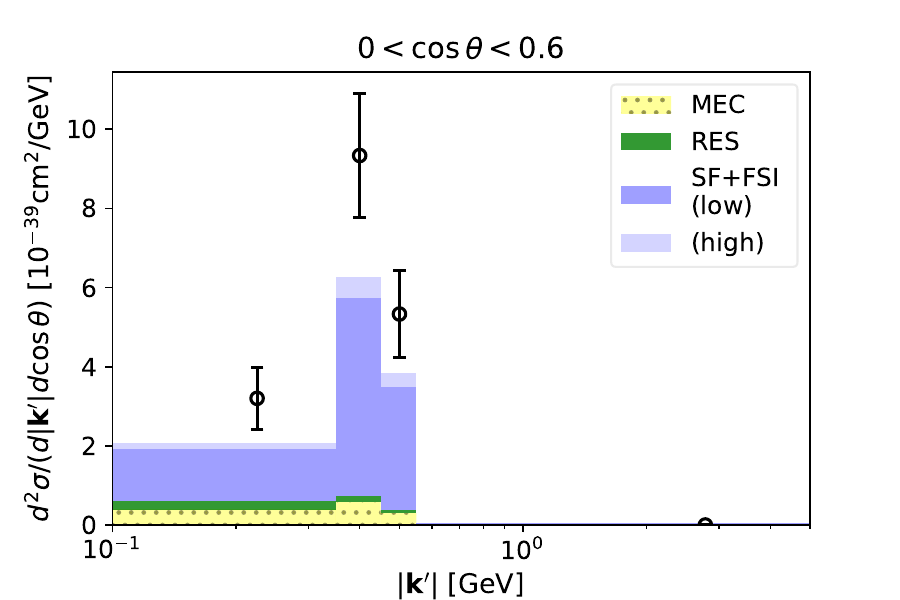}

  \caption{The same as Fig.~\ref{fig:t2k}, showing theoretical uncertainty coming from the reconstruction of spectral function. ``SF+FSI (low)'' was obtained using the lower bound of the spectral reconstruction, while ``SF+FSI (high)'' corresponds to the upper bound.} 
  \label{fig:t2k_err}
\end{figure*}

The results of the simulation for the double differential cross section $\nu_\mu + ^{16}\mathrm{O} \to \mu^- + X$ of CC0$\pi$ events done with NuWro and our SF are presented in Fig.~\ref{fig:t2k}. In the same plot we also show the predictions for the SF with the inclusion of optical potential (hatched pattern, denoted with SF+FSI). Here, we do not account for the theoretical uncertainty of spectral function, focusing only on the role played by optical potential. Other dynamical channels, MEC and resonance contributions (RES), were chosen to be the same as explained in Ref.~\cite{T2K:2020jav}.
We are aware that these predictions are not fully consistent, since the theoretical description of each mechanism is based on a different model. To make our comparison with the experimental data more meaningful, we would need to address all the contributions within the same SF method. This is certainly an important direction of future investigations. 
However, at this point we  focus only on the IA mechanism.
In Fig.~\ref{fig:t2k_err} we show our final prediction, including the uncertainty of SF. It leads up to $\sim 10\%$ effect, depending on the considered kinematics. 
We find a reasonable agreement with the data, similar to the results of Ref.~\cite{T2K:2020jav} where several MC event generators were employed (a variety of models were used for the QE mechanism, including random phase approximation corrections, a phenomenological spectral function or relativistic mean field calculations). Our SF quenches the response when compared to the local Fermi Gas, even by $25\%$ for forward scattering angles.
In fact, the kinematics of the most forward angles (upper left panel in Fig.~\ref{fig:t2k}) depend mostly on the details of the nuclear model, since the momentum transfer is the smallest (covers mostly the range of $|\mathbf{q}| \approx 100-300$ MeV). 
For the angles $0.96 < \cos\theta < 1$ the optical potential causes a substantial depletion of the bins corresponding to the values of $|\mathbf{k}^\prime|<800$ MeV. 
Our simple model of FSI gives reasonable results for the electron scattering at the intermediate momentum transfer $|\mathbf{q}|\gtrapprox 450$ MeV shifting the QE peak to lower energies (see Fig.~\ref{fig:o16_electron_scattering}). Here, on the contrary, we observe a significant effect. At this kinematics (low $|\mathbf{q}|$ and forward scattering angles) the IA is much less reliable. 
This region of phase-space escapes the capability of our method and should be rather described by a consistent calculation of FSI, available with the LIT-CC approach. We also observe that the contribution coming from the phenomenological MEC is substantial in this range. This prediction can also be verified using an ab-initio approach including one- and two-body currents~\cite{Lovato_2020, Pastore:2019urn}.

\section{Conclusion and outlook}
\label{sec:conclusion}
We calculated spectral functions of $^{16}$O within the many-body coupled-cluster framework and employing a chiral nuclear Hamiltonian including 3N forces at next-to-next-to leading order. The procedure required a reconstruction of the spectral properties (i.e. the energy-dependant part of the SF), which we performed within the ChEK method. This approach, which was benchmarked on the $^4$He in an earlier publication~\cite{Sobczyk:2022ezo}, allows to assess the uncertainty of our calculation and to propagate it to the cross-section results.

Within the impulse approximation, the SF can be directly related to the scattering cross-section. Using this assumption, we give predictions for the lepton-nucleus scattering in the QE regime both for electron and neutrino scattering.
The electron scattering data for $^{16}$O are scarce and cover only the medium and high momentum transfer regions.
 
Within this energy range the IA works well and we get a good agreement with the data, although for $|\mathbf{q}|<500$ MeV the FSI play an important role and the inclusion of optical potential visibly improves the agreement. We still do not account for the absorption of the outgoing nucleon, which is certainly an important topic to be addressed in the future when aiming at the comparison with inclusive data.
We would like to point out that further investigations of the QE region are currently restricted due to the lack of low-energy electron-scattering data on $^{16}$O.
More data would be of great value to guide theoretical models used in the future T2HK experiment. There are plans to take new data  on $^{16}$O in the future at MAMI in Germany~\cite{whitepaper1}.

We presented a comparison with neutrino T2K data for CC$0\pi$ events which are sensitive to the QE mechanism. To this end, we implemented our SF in the NuWro MC generator. In our analysis we observed that for the forward angles the optical potential plays an important role. In fact, the IA picture becomes much less reliable in this regime and a consistent calculation which accounts for the final state interactions, as the LIT-CC, would be more appropriate. Also the role played by the two-body currents should be examined. The work in this direction is already on-going.
Since our studies are mainly motivated by the neutrino oscillation experiments, we find it important to make our spectral functions available for further exploration~\cite{repo}.

\begin{acknowledgements}
We acknowledge useful discussions with G. Hagen and T. Papenbrock and we thank them for letting us use the NuCCore coupled-cluster code.
J.E.S. acknowledges the support of the Humboldt Foundation through a Humboldt Research Fellowship for Postdoctoral Researchers.
This project has received funding from the European Union’s Horizon 2020 research and innovation programme under the Marie Skłodowska-Curie grant agreement No. 101026014.
This work was supported in part by the Deutsche
Forschungsgemeinschaft (DFG)
through the Cluster of Excellence ``Precision Physics, Fundamental
Interactions, and Structure of Matter" (PRISMA$^+$ EXC 2118/1) funded by the
DFG within the German Excellence Strategy (Project ID 39083149)
\end{acknowledgements}

\bibliography{biblio}

\begin{thebibliography}{41}%
\makeatletter
\providecommand \@ifxundefined [1]{%
 \@ifx{#1\undefined}
}%
\providecommand \@ifnum [1]{%
 \ifnum #1\expandafter \@firstoftwo
 \else \expandafter \@secondoftwo
 \fi
}%
\providecommand \@ifx [1]{%
 \ifx #1\expandafter \@firstoftwo
 \else \expandafter \@secondoftwo
 \fi
}%
\providecommand \natexlab [1]{#1}%
\providecommand \enquote  [1]{``#1''}%
\providecommand \bibnamefont  [1]{#1}%
\providecommand \bibfnamefont [1]{#1}%
\providecommand \citenamefont [1]{#1}%
\providecommand \href@noop [0]{\@secondoftwo}%
\providecommand \href [0]{\begingroup \@sanitize@url \@href}%
\providecommand \@href[1]{\@@startlink{#1}\@@href}%
\providecommand \@@href[1]{\endgroup#1\@@endlink}%
\providecommand \@sanitize@url [0]{\catcode `\\12\catcode `\$12\catcode
  `\&12\catcode `\#12\catcode `\^12\catcode `\_12\catcode `\%12\relax}%
\providecommand \@@startlink[1]{}%
\providecommand \@@endlink[0]{}%
\providecommand \url  [0]{\begingroup\@sanitize@url \@url }%
\providecommand \@url [1]{\endgroup\@href {#1}{\urlprefix }}%
\providecommand \urlprefix  [0]{URL }%
\providecommand \Eprint [0]{\href }%
\providecommand \doibase [0]{http://dx.doi.org/}%
\providecommand \selectlanguage [0]{\@gobble}%
\providecommand \bibinfo  [0]{\@secondoftwo}%
\providecommand \bibfield  [0]{\@secondoftwo}%
\providecommand \translation [1]{[#1]}%
\providecommand \BibitemOpen [0]{}%
\providecommand \bibitemStop [0]{}%
\providecommand \bibitemNoStop [0]{.\EOS\space}%
\providecommand \EOS [0]{\spacefactor3000\relax}%
\providecommand \BibitemShut  [1]{\csname bibitem#1\endcsname}%
\let\auto@bib@innerbib\@empty
\bibitem [{\citenamefont {Acciarri}\ \emph {et~al.}(2015)\citenamefont
  {Acciarri} \emph {et~al.}}]{DUNE}%
  \BibitemOpen
  \bibfield  {author} {\bibinfo {author} {\bibfnamefont {R.}~\bibnamefont
  {Acciarri}} \emph {et~al.} (\bibinfo {collaboration} {DUNE}),\ }\bibfield
  {title} {\enquote {\bibinfo {title} {{Long-Baseline Neutrino Facility (LBNF)
  and Deep Underground Neutrino Experiment (DUNE)}},}\ }\href@noop {} {\
  (\bibinfo {year} {2015})},\ \Eprint {http://arxiv.org/abs/1512.06148}
  {arXiv:1512.06148 [physics.ins-det]} \BibitemShut {NoStop}%
\bibitem [{\citenamefont {Abe}\ \emph {et~al.}(2015)\citenamefont {Abe} \emph
  {et~al.}}]{hyperk}%
  \BibitemOpen
  \bibfield  {author} {\bibinfo {author} {\bibfnamefont {K.}~\bibnamefont
  {Abe}} \emph {et~al.} (\bibinfo {collaboration} {Hyper-Kamiokande
  Proto-Collaboration}),\ }\bibfield  {title} {\enquote {\bibinfo {title}
  {{Physics potential of a long-baseline neutrino oscillation experiment using
  a J-PARC neutrino beam and Hyper-Kamiokande}},}\ }\href {\doibase
  10.1093/ptep/ptv061} {\bibfield  {journal} {\bibinfo  {journal} {PTEP}\
  }\textbf {\bibinfo {volume} {2015}},\ \bibinfo {pages} {053C02} (\bibinfo
  {year} {2015})}\BibitemShut {NoStop}%
\bibitem [{\citenamefont {Alvarez-Ruso}\ \emph {et~al.}(2018)\citenamefont
  {Alvarez-Ruso} \emph {et~al.}}]{Nustec}%
  \BibitemOpen
  \bibfield  {author} {\bibinfo {author} {\bibfnamefont {L.}~\bibnamefont
  {Alvarez-Ruso}} \emph {et~al.} (\bibinfo {collaboration} {NuSTEC}),\
  }\bibfield  {title} {\enquote {\bibinfo {title} {{NuSTEC White Paper: Status
  and challenges of neutrino\textendash{}nucleus scattering}},}\ }\href
  {\doibase 10.1016/j.ppnp.2018.01.006} {\bibfield  {journal} {\bibinfo
  {journal} {Prog. Part. Nucl. Phys.}\ }\textbf {\bibinfo {volume} {100}},\
  \bibinfo {pages} {1--68} (\bibinfo {year} {2018})},\ \Eprint
  {http://arxiv.org/abs/1706.03621} {arXiv:1706.03621 [hep-ph]} \BibitemShut
  {NoStop}%
\bibitem [{\citenamefont {Ankowski}\ \emph {et~al.}(2022)\citenamefont
  {Ankowski} \emph {et~al.}}]{whitepaper1}%
  \BibitemOpen
  \bibfield  {author} {\bibinfo {author} {\bibfnamefont {A.~M.}\ \bibnamefont
  {Ankowski}} \emph {et~al.},\ }\bibfield  {title} {\enquote {\bibinfo {title}
  {{Electron Scattering and Neutrino Physics}},}\ }\href@noop {} {\  (\bibinfo
  {year} {2022})},\ \Eprint {http://arxiv.org/abs/2203.06853} {arXiv:2203.06853
  [hep-ex]} \BibitemShut {NoStop}%
\bibitem [{\citenamefont {Ruso}\ \emph {et~al.}(2022)\citenamefont {Ruso} \emph
  {et~al.}}]{whitepaper2}%
  \BibitemOpen
  \bibfield  {author} {\bibinfo {author} {\bibfnamefont {L.~Alvarez}\
  \bibnamefont {Ruso}} \emph {et~al.},\ }\bibfield  {title} {\enquote {\bibinfo
  {title} {{Theoretical tools for neutrino scattering: interplay between
  lattice QCD, EFTs, nuclear physics, phenomenology, and neutrino event
  generators}},}\ }\href@noop {} {\  (\bibinfo {year} {2022})},\ \Eprint
  {http://arxiv.org/abs/2203.09030} {arXiv:2203.09030 [hep-ph]} \BibitemShut
  {NoStop}%
\bibitem [{\citenamefont {Lovato}\ \emph {et~al.}(2018)\citenamefont {Lovato},
  \citenamefont {Gandolfi}, \citenamefont {Carlson}, \citenamefont {Lusk},
  \citenamefont {Pieper},\ and\ \citenamefont {Schiavilla}}]{Lovato:2017cux}%
  \BibitemOpen
  \bibfield  {author} {\bibinfo {author} {\bibfnamefont {A.}~\bibnamefont
  {Lovato}}, \bibinfo {author} {\bibfnamefont {S.}~\bibnamefont {Gandolfi}},
  \bibinfo {author} {\bibfnamefont {J.}~\bibnamefont {Carlson}}, \bibinfo
  {author} {\bibfnamefont {Ewing}\ \bibnamefont {Lusk}}, \bibinfo {author}
  {\bibfnamefont {Steven~C.}\ \bibnamefont {Pieper}}, \ and\ \bibinfo {author}
  {\bibfnamefont {R.}~\bibnamefont {Schiavilla}},\ }\bibfield  {title}
  {\enquote {\bibinfo {title} {{Quantum Monte Carlo calculation of
  neutral-current $\nu-^{12}C$ inclusive quasielastic scattering}},}\ }\href
  {\doibase 10.1103/PhysRevC.97.022502} {\bibfield  {journal} {\bibinfo
  {journal} {Phys. Rev. C}\ }\textbf {\bibinfo {volume} {97}},\ \bibinfo
  {pages} {022502} (\bibinfo {year} {2018})},\ \Eprint
  {http://arxiv.org/abs/1711.02047} {arXiv:1711.02047 [nucl-th]} \BibitemShut
  {NoStop}%
\bibitem [{\citenamefont {Lovato}\ \emph {et~al.}(2020)\citenamefont {Lovato},
  \citenamefont {Carlson}, \citenamefont {Gandolfi}, \citenamefont {Rocco},\
  and\ \citenamefont {Schiavilla}}]{Lovato_2020}%
  \BibitemOpen
  \bibfield  {author} {\bibinfo {author} {\bibfnamefont {A.}~\bibnamefont
  {Lovato}}, \bibinfo {author} {\bibfnamefont {J.}~\bibnamefont {Carlson}},
  \bibinfo {author} {\bibfnamefont {S.}~\bibnamefont {Gandolfi}}, \bibinfo
  {author} {\bibfnamefont {N.}~\bibnamefont {Rocco}}, \ and\ \bibinfo {author}
  {\bibfnamefont {R.}~\bibnamefont {Schiavilla}},\ }\bibfield  {title}
  {\enquote {\bibinfo {title} {Ab initio study of
  $({\ensuremath{\nu}}_{\ensuremath{\ell}},{\ensuremath{\ell}}^{\ensuremath{-}})$
  and
  $({\overline{\ensuremath{\nu}}}_{\ensuremath{\ell}},{\ensuremath{\ell}}^{+})$
  inclusive scattering in $^{12}\mathrm{C}$: Confronting the miniboone and t2k
  ccqe data},}\ }\href {\doibase 10.1103/PhysRevX.10.031068} {\bibfield
  {journal} {\bibinfo  {journal} {Phys. Rev. X}\ }\textbf {\bibinfo {volume}
  {10}},\ \bibinfo {pages} {031068} (\bibinfo {year} {2020})}\BibitemShut
  {NoStop}%
\bibitem [{\citenamefont {Sobczyk}\ \emph {et~al.}(2020)\citenamefont
  {Sobczyk}, \citenamefont {Acharya}, \citenamefont {Bacca},\ and\
  \citenamefont {Hagen}}]{Sobczyk:2020qtw}%
  \BibitemOpen
  \bibfield  {author} {\bibinfo {author} {\bibfnamefont {J.~E.}\ \bibnamefont
  {Sobczyk}}, \bibinfo {author} {\bibfnamefont {B.}~\bibnamefont {Acharya}},
  \bibinfo {author} {\bibfnamefont {S.}~\bibnamefont {Bacca}}, \ and\ \bibinfo
  {author} {\bibfnamefont {G.}~\bibnamefont {Hagen}},\ }\bibfield  {title}
  {\enquote {\bibinfo {title} {{Coulomb sum rule for $^4$He and $^{16}$O from
  coupled-cluster theory}},}\ }\href {\doibase 10.1103/PhysRevC.102.064312}
  {\bibfield  {journal} {\bibinfo  {journal} {Phys. Rev. C}\ }\textbf {\bibinfo
  {volume} {102}},\ \bibinfo {pages} {064312} (\bibinfo {year} {2020})},\
  \Eprint {http://arxiv.org/abs/2009.01761} {arXiv:2009.01761 [nucl-th]}
  \BibitemShut {NoStop}%
\bibitem [{\citenamefont {Sobczyk}\ \emph {et~al.}(2021)\citenamefont
  {Sobczyk}, \citenamefont {Acharya}, \citenamefont {Bacca},\ and\
  \citenamefont {Hagen}}]{Sobczyk:2021dwm}%
  \BibitemOpen
  \bibfield  {author} {\bibinfo {author} {\bibfnamefont {J.~E.}\ \bibnamefont
  {Sobczyk}}, \bibinfo {author} {\bibfnamefont {B.}~\bibnamefont {Acharya}},
  \bibinfo {author} {\bibfnamefont {S.}~\bibnamefont {Bacca}}, \ and\ \bibinfo
  {author} {\bibfnamefont {G.}~\bibnamefont {Hagen}},\ }\bibfield  {title}
  {\enquote {\bibinfo {title} {{Ab initio computation of the longitudinal
  response function in $^{40}$Ca}},}\ }\href {\doibase
  10.1103/PhysRevLett.127.072501} {\bibfield  {journal} {\bibinfo  {journal}
  {Phys. Rev. Lett.}\ }\textbf {\bibinfo {volume} {127}},\ \bibinfo {pages}
  {072501} (\bibinfo {year} {2021})},\ \Eprint
  {http://arxiv.org/abs/2103.06786} {arXiv:2103.06786 [nucl-th]} \BibitemShut
  {NoStop}%
\bibitem [{\citenamefont {Rocco}\ \emph {et~al.}(2019)\citenamefont {Rocco},
  \citenamefont {Nakamura}, \citenamefont {Lee},\ and\ \citenamefont
  {Lovato}}]{Rocco:2019gfb}%
  \BibitemOpen
  \bibfield  {author} {\bibinfo {author} {\bibfnamefont {Noemi}\ \bibnamefont
  {Rocco}}, \bibinfo {author} {\bibfnamefont {Satoshi~X.}\ \bibnamefont
  {Nakamura}}, \bibinfo {author} {\bibfnamefont {T.~S.~H.}\ \bibnamefont
  {Lee}}, \ and\ \bibinfo {author} {\bibfnamefont {Alessandro}\ \bibnamefont
  {Lovato}},\ }\bibfield  {title} {\enquote {\bibinfo {title} {{Electroweak
  Pion-Production on Nuclei within the Extended Factorization Scheme}},}\
  }\href {\doibase 10.1103/PhysRevC.100.045503} {\bibfield  {journal} {\bibinfo
   {journal} {Phys. Rev. C}\ }\textbf {\bibinfo {volume} {100}},\ \bibinfo
  {pages} {045503} (\bibinfo {year} {2019})},\ \Eprint
  {http://arxiv.org/abs/1907.01093} {arXiv:1907.01093 [nucl-th]} \BibitemShut
  {NoStop}%
\bibitem [{\citenamefont {Jiang}\ \emph {et~al.}(2022)\citenamefont {Jiang}
  \emph {et~al.}}]{JeffersonLabHallA:2022cit}%
  \BibitemOpen
  \bibfield  {author} {\bibinfo {author} {\bibfnamefont {L.}~\bibnamefont
  {Jiang}} \emph {et~al.} (\bibinfo {collaboration} {Jefferson Lab Hall A}),\
  }\bibfield  {title} {\enquote {\bibinfo {title} {{Determination of the argon
  spectral function from (e,e'p) data}},}\ }\href {\doibase
  10.1103/PhysRevD.105.112002} {\bibfield  {journal} {\bibinfo  {journal}
  {Phys. Rev. D}\ }\textbf {\bibinfo {volume} {105}},\ \bibinfo {pages}
  {112002} (\bibinfo {year} {2022})},\ \Eprint
  {http://arxiv.org/abs/2203.01748} {arXiv:2203.01748 [nucl-ex]} \BibitemShut
  {NoStop}%
\bibitem [{\citenamefont {Benhar}\ \emph {et~al.}(1994)\citenamefont {Benhar},
  \citenamefont {Fabrocini}, \citenamefont {Fantoni},\ and\ \citenamefont
  {Sick}}]{Benhar:1994hw}%
  \BibitemOpen
  \bibfield  {author} {\bibinfo {author} {\bibfnamefont {O.}~\bibnamefont
  {Benhar}}, \bibinfo {author} {\bibfnamefont {A.}~\bibnamefont {Fabrocini}},
  \bibinfo {author} {\bibfnamefont {S.}~\bibnamefont {Fantoni}}, \ and\
  \bibinfo {author} {\bibfnamefont {I.}~\bibnamefont {Sick}},\ }\bibfield
  {title} {\enquote {\bibinfo {title} {{Spectral function of finite nuclei and
  scattering of GeV electrons}},}\ }\href {\doibase
  10.1016/0375-9474(94)90920-2} {\bibfield  {journal} {\bibinfo  {journal}
  {Nucl. Phys. A}\ }\textbf {\bibinfo {volume} {579}},\ \bibinfo {pages}
  {493--517} (\bibinfo {year} {1994})}\BibitemShut {NoStop}%
\bibitem [{\citenamefont {Nieves}\ and\ \citenamefont
  {Sobczyk}(2017)}]{Nieves:2017lij}%
  \BibitemOpen
  \bibfield  {author} {\bibinfo {author} {\bibfnamefont {Juan}\ \bibnamefont
  {Nieves}}\ and\ \bibinfo {author} {\bibfnamefont {Joanna~Ewa}\ \bibnamefont
  {Sobczyk}},\ }\bibfield  {title} {\enquote {\bibinfo {title} {{In medium
  dispersion relation effects in nuclear inclusive reactions at intermediate
  and low energies}},}\ }\href {\doibase 10.1016/j.aop.2017.06.002} {\bibfield
  {journal} {\bibinfo  {journal} {Annals Phys.}\ }\textbf {\bibinfo {volume}
  {383}},\ \bibinfo {pages} {455--496} (\bibinfo {year} {2017})},\ \Eprint
  {http://arxiv.org/abs/1701.03628} {arXiv:1701.03628 [nucl-th]} \BibitemShut
  {NoStop}%
\bibitem [{\citenamefont {Buss}\ \emph {et~al.}(2007)\citenamefont {Buss},
  \citenamefont {Leitner}, \citenamefont {Mosel},\ and\ \citenamefont
  {Alvarez-Ruso}}]{Buss:2007ar}%
  \BibitemOpen
  \bibfield  {author} {\bibinfo {author} {\bibfnamefont {O.}~\bibnamefont
  {Buss}}, \bibinfo {author} {\bibfnamefont {T.}~\bibnamefont {Leitner}},
  \bibinfo {author} {\bibfnamefont {U.}~\bibnamefont {Mosel}}, \ and\ \bibinfo
  {author} {\bibfnamefont {L.}~\bibnamefont {Alvarez-Ruso}},\ }\bibfield
  {title} {\enquote {\bibinfo {title} {{The Influence of the nuclear medium on
  inclusive electron and neutrino scattering off nuclei}},}\ }\href {\doibase
  10.1103/PhysRevC.76.035502} {\bibfield  {journal} {\bibinfo  {journal} {Phys.
  Rev. C}\ }\textbf {\bibinfo {volume} {76}},\ \bibinfo {pages} {035502}
  (\bibinfo {year} {2007})},\ \Eprint {http://arxiv.org/abs/0707.0232}
  {arXiv:0707.0232 [nucl-th]} \BibitemShut {NoStop}%
\bibitem [{\citenamefont {Rocco}\ and\ \citenamefont
  {Barbieri}(2018)}]{Rocco:2018vbf}%
  \BibitemOpen
  \bibfield  {author} {\bibinfo {author} {\bibfnamefont {N.}~\bibnamefont
  {Rocco}}\ and\ \bibinfo {author} {\bibfnamefont {C.}~\bibnamefont
  {Barbieri}},\ }\bibfield  {title} {\enquote {\bibinfo {title} {{Inclusive
  electron-nucleus cross section within the Self Consistent Green's Function
  approach}},}\ }\href {\doibase 10.1103/PhysRevC.98.025501} {\bibfield
  {journal} {\bibinfo  {journal} {Phys. Rev. C}\ }\textbf {\bibinfo {volume}
  {98}},\ \bibinfo {pages} {025501} (\bibinfo {year} {2018})},\ \Eprint
  {http://arxiv.org/abs/1803.00825} {arXiv:1803.00825 [nucl-th]} \BibitemShut
  {NoStop}%
\bibitem [{\citenamefont {Barbieri}\ \emph {et~al.}(2019)\citenamefont
  {Barbieri}, \citenamefont {Rocco},\ and\ \citenamefont
  {Som\`a}}]{Barbieri:2019ual}%
  \BibitemOpen
  \bibfield  {author} {\bibinfo {author} {\bibfnamefont {C.}~\bibnamefont
  {Barbieri}}, \bibinfo {author} {\bibfnamefont {N.}~\bibnamefont {Rocco}}, \
  and\ \bibinfo {author} {\bibfnamefont {V.}~\bibnamefont {Som\`a}},\
  }\bibfield  {title} {\enquote {\bibinfo {title} {{Lepton Scattering from
  $^{40}$Ar and Ti in the Quasielastic Peak Region}},}\ }\href {\doibase
  10.1103/PhysRevC.100.062501} {\bibfield  {journal} {\bibinfo  {journal}
  {Phys. Rev. C}\ }\textbf {\bibinfo {volume} {100}},\ \bibinfo {pages}
  {062501} (\bibinfo {year} {2019})},\ \Eprint
  {http://arxiv.org/abs/1907.01122} {arXiv:1907.01122 [nucl-th]} \BibitemShut
  {NoStop}%
\bibitem [{\citenamefont {Roggero}(2020)}]{Roggero:2020qoz}%
  \BibitemOpen
  \bibfield  {author} {\bibinfo {author} {\bibfnamefont {Alessandro}\
  \bibnamefont {Roggero}},\ }\bibfield  {title} {\enquote {\bibinfo {title}
  {{Spectral density estimation with the Gaussian Integral Transform}},}\
  }\href {\doibase 10.1103/PhysRevA.102.022409} {\bibfield  {journal} {\bibinfo
   {journal} {Phys. Rev. A}\ }\textbf {\bibinfo {volume} {102}},\ \bibinfo
  {pages} {022409} (\bibinfo {year} {2020})},\ \Eprint
  {http://arxiv.org/abs/2004.04889} {arXiv:2004.04889 [quant-ph]} \BibitemShut
  {NoStop}%
\bibitem [{\citenamefont {Sobczyk}\ and\ \citenamefont
  {Roggero}(2021)}]{Sobczyk:2021ejs}%
  \BibitemOpen
  \bibfield  {author} {\bibinfo {author} {\bibfnamefont {Joanna~E.}\
  \bibnamefont {Sobczyk}}\ and\ \bibinfo {author} {\bibfnamefont {Alessandro}\
  \bibnamefont {Roggero}},\ }\bibfield  {title} {\enquote {\bibinfo {title}
  {{Spectral density reconstruction with Chebyshev polynomials}},}\ }\href@noop
  {} {\  (\bibinfo {year} {2021})},\ \Eprint {http://arxiv.org/abs/2110.02108}
  {arXiv:2110.02108 [nucl-th]} \BibitemShut {NoStop}%
\bibitem [{\citenamefont {Hagen}\ \emph {et~al.}(2014)\citenamefont {Hagen},
  \citenamefont {Papenbrock}, \citenamefont {Hjorth-Jensen},\ and\
  \citenamefont {Dean}}]{hagen2014}%
  \BibitemOpen
  \bibfield  {author} {\bibinfo {author} {\bibfnamefont {G.}~\bibnamefont
  {Hagen}}, \bibinfo {author} {\bibfnamefont {T.}~\bibnamefont {Papenbrock}},
  \bibinfo {author} {\bibfnamefont {M.}~\bibnamefont {Hjorth-Jensen}}, \ and\
  \bibinfo {author} {\bibfnamefont {D.~J.}\ \bibnamefont {Dean}},\ }\bibfield
  {title} {\enquote {\bibinfo {title} {Coupled-cluster computations of atomic
  nuclei},}\ }\href {\doibase 10.1088/0034-4885/77/9/096302} {\bibfield
  {journal} {\bibinfo  {journal} {Rep. Prog. Phys.}\ }\textbf {\bibinfo
  {volume} {77}},\ \bibinfo {pages} {096302} (\bibinfo {year}
  {2014})}\BibitemShut {NoStop}%
\bibitem [{\citenamefont {Sobczyk}\ \emph {et~al.}(2022)\citenamefont
  {Sobczyk}, \citenamefont {Bacca}, \citenamefont {Hagen},\ and\ \citenamefont
  {Papenbrock}}]{Sobczyk:2022ezo}%
  \BibitemOpen
  \bibfield  {author} {\bibinfo {author} {\bibfnamefont {J.~E.}\ \bibnamefont
  {Sobczyk}}, \bibinfo {author} {\bibfnamefont {S.}~\bibnamefont {Bacca}},
  \bibinfo {author} {\bibfnamefont {G.}~\bibnamefont {Hagen}}, \ and\ \bibinfo
  {author} {\bibfnamefont {T.}~\bibnamefont {Papenbrock}},\ }\bibfield  {title}
  {\enquote {\bibinfo {title} {{Spectral function for $^4$He using the
  Chebyshev expansion in coupled-cluster theory}},}\ }\href@noop {} {\
  (\bibinfo {year} {2022})},\ \Eprint {http://arxiv.org/abs/2205.03592}
  {arXiv:2205.03592 [nucl-th]} \BibitemShut {NoStop}%
\bibitem [{\citenamefont {Juszczak}\ \emph {et~al.}(2006)\citenamefont
  {Juszczak}, \citenamefont {Nowak},\ and\ \citenamefont
  {Sobczyk}}]{Juszczak:2005zs}%
  \BibitemOpen
  \bibfield  {author} {\bibinfo {author} {\bibfnamefont {Cezary}\ \bibnamefont
  {Juszczak}}, \bibinfo {author} {\bibfnamefont {Jaroslaw~A.}\ \bibnamefont
  {Nowak}}, \ and\ \bibinfo {author} {\bibfnamefont {Jan~T.}\ \bibnamefont
  {Sobczyk}},\ }\bibfield  {title} {\enquote {\bibinfo {title} {{Simulations
  from a new neutrino event generator}},}\ }\href {\doibase
  10.1016/j.nuclphysbps.2006.08.069} {\bibfield  {journal} {\bibinfo  {journal}
  {Nucl. Phys. B Proc. Suppl.}\ }\textbf {\bibinfo {volume} {159}},\ \bibinfo
  {pages} {211--216} (\bibinfo {year} {2006})},\ \Eprint
  {http://arxiv.org/abs/hep-ph/0512365} {arXiv:hep-ph/0512365} \BibitemShut
  {NoStop}%
\bibitem [{\citenamefont {Golan}\ \emph {et~al.}(2012)\citenamefont {Golan},
  \citenamefont {Sobczyk},\ and\ \citenamefont {Zmuda}}]{Golan:2012rfa}%
  \BibitemOpen
  \bibfield  {author} {\bibinfo {author} {\bibfnamefont {T.}~\bibnamefont
  {Golan}}, \bibinfo {author} {\bibfnamefont {J.~T.}\ \bibnamefont {Sobczyk}},
  \ and\ \bibinfo {author} {\bibfnamefont {J.}~\bibnamefont {Zmuda}},\
  }\bibfield  {title} {\enquote {\bibinfo {title} {{NuWro: the Wroclaw Monte
  Carlo Generator of Neutrino Interactions}},}\ }\href {\doibase
  10.1016/j.nuclphysbps.2012.09.136} {\bibfield  {journal} {\bibinfo  {journal}
  {Nucl. Phys. B Proc. Suppl.}\ }\textbf {\bibinfo {volume} {229-232}},\
  \bibinfo {pages} {499--499} (\bibinfo {year} {2012})}\BibitemShut {NoStop}%
\bibitem [{\citenamefont {Bradford}\ \emph {et~al.}(2006)\citenamefont
  {Bradford}, \citenamefont {Bodek}, \citenamefont {Budd},\ and\ \citenamefont
  {Arrington}}]{Bradford:2006yz}%
  \BibitemOpen
  \bibfield  {author} {\bibinfo {author} {\bibfnamefont {R.}~\bibnamefont
  {Bradford}}, \bibinfo {author} {\bibfnamefont {A.}~\bibnamefont {Bodek}},
  \bibinfo {author} {\bibfnamefont {Howard~Scott}\ \bibnamefont {Budd}}, \ and\
  \bibinfo {author} {\bibfnamefont {J.}~\bibnamefont {Arrington}},\ }\bibfield
  {title} {\enquote {\bibinfo {title} {{A New parameterization of the nucleon
  elastic form-factors}},}\ }\href {\doibase 10.1016/j.nuclphysbps.2006.08.028}
  {\bibfield  {journal} {\bibinfo  {journal} {Nucl. Phys. B Proc. Suppl.}\
  }\textbf {\bibinfo {volume} {159}},\ \bibinfo {pages} {127--132} (\bibinfo
  {year} {2006})},\ \Eprint {http://arxiv.org/abs/hep-ex/0602017}
  {arXiv:hep-ex/0602017} \BibitemShut {NoStop}%
\bibitem [{\citenamefont {Sobczyk}(2017)}]{Sobczyk:2017mts}%
  \BibitemOpen
  \bibfield  {author} {\bibinfo {author} {\bibfnamefont {Joanna~Ewa}\
  \bibnamefont {Sobczyk}},\ }\bibfield  {title} {\enquote {\bibinfo {title}
  {{Intercomparison of lepton-nucleus scattering models in the quasielastic
  region}},}\ }\href {\doibase 10.1103/PhysRevC.96.045501} {\bibfield
  {journal} {\bibinfo  {journal} {Phys. Rev. C}\ }\textbf {\bibinfo {volume}
  {96}},\ \bibinfo {pages} {045501} (\bibinfo {year} {2017})},\ \Eprint
  {http://arxiv.org/abs/1706.06739} {arXiv:1706.06739 [nucl-th]} \BibitemShut
  {NoStop}%
\bibitem [{\citenamefont {Cooper}\ \emph {et~al.}(1993)\citenamefont {Cooper},
  \citenamefont {Hama}, \citenamefont {Clark},\ and\ \citenamefont
  {Mercer}}]{Cooper:1993nx}%
  \BibitemOpen
  \bibfield  {author} {\bibinfo {author} {\bibfnamefont {E.~D.}\ \bibnamefont
  {Cooper}}, \bibinfo {author} {\bibfnamefont {S.}~\bibnamefont {Hama}},
  \bibinfo {author} {\bibfnamefont {B.~C.}\ \bibnamefont {Clark}}, \ and\
  \bibinfo {author} {\bibfnamefont {R.~L.}\ \bibnamefont {Mercer}},\ }\bibfield
   {title} {\enquote {\bibinfo {title} {{Global Dirac phenomenology for proton
  nucleus elastic scattering}},}\ }\href {\doibase 10.1103/PhysRevC.47.297}
  {\bibfield  {journal} {\bibinfo  {journal} {Phys. Rev. C}\ }\textbf {\bibinfo
  {volume} {47}},\ \bibinfo {pages} {297--311} (\bibinfo {year}
  {1993})}\BibitemShut {NoStop}%
\bibitem [{\citenamefont {Ankowski}\ \emph {et~al.}(2015)\citenamefont
  {Ankowski}, \citenamefont {Benhar},\ and\ \citenamefont
  {Sakuda}}]{Ankowski:2014yfa}%
  \BibitemOpen
  \bibfield  {author} {\bibinfo {author} {\bibfnamefont {Artur~M.}\
  \bibnamefont {Ankowski}}, \bibinfo {author} {\bibfnamefont {Omar}\
  \bibnamefont {Benhar}}, \ and\ \bibinfo {author} {\bibfnamefont {Makoto}\
  \bibnamefont {Sakuda}},\ }\bibfield  {title} {\enquote {\bibinfo {title}
  {{Improving the accuracy of neutrino energy reconstruction in charged-current
  quasielastic scattering off nuclear targets}},}\ }\href {\doibase
  10.1103/PhysRevD.91.033005} {\bibfield  {journal} {\bibinfo  {journal} {Phys.
  Rev. D}\ }\textbf {\bibinfo {volume} {91}},\ \bibinfo {pages} {033005}
  (\bibinfo {year} {2015})},\ \Eprint {http://arxiv.org/abs/1404.5687}
  {arXiv:1404.5687 [nucl-th]} \BibitemShut {NoStop}%
\bibitem [{\citenamefont {Benhar}\ \emph {et~al.}(1991)\citenamefont {Benhar},
  \citenamefont {Fabrocini}, \citenamefont {Fantoni}, \citenamefont {Miller},
  \citenamefont {Pandharipande},\ and\ \citenamefont {Sick}}]{Benhar:1991af}%
  \BibitemOpen
  \bibfield  {author} {\bibinfo {author} {\bibfnamefont {O.}~\bibnamefont
  {Benhar}}, \bibinfo {author} {\bibfnamefont {A.}~\bibnamefont {Fabrocini}},
  \bibinfo {author} {\bibfnamefont {S.}~\bibnamefont {Fantoni}}, \bibinfo
  {author} {\bibfnamefont {G.~A.}\ \bibnamefont {Miller}}, \bibinfo {author}
  {\bibfnamefont {V.~R.}\ \bibnamefont {Pandharipande}}, \ and\ \bibinfo
  {author} {\bibfnamefont {I.}~\bibnamefont {Sick}},\ }\bibfield  {title}
  {\enquote {\bibinfo {title} {{Scattering of GeV electrons by nuclear
  matter}},}\ }\href {\doibase 10.1103/PhysRevC.44.2328} {\bibfield  {journal}
  {\bibinfo  {journal} {Phys. Rev. C}\ }\textbf {\bibinfo {volume} {44}},\
  \bibinfo {pages} {2328--2342} (\bibinfo {year} {1991})}\BibitemShut {NoStop}%
\bibitem [{\citenamefont {Efros}\ \emph {et~al.}(1994)\citenamefont {Efros},
  \citenamefont {Leidemann},\ and\ \citenamefont {Orlandini}}]{Efros:1994iq}%
  \BibitemOpen
  \bibfield  {author} {\bibinfo {author} {\bibfnamefont {Victor~D.}\
  \bibnamefont {Efros}}, \bibinfo {author} {\bibfnamefont {Winfried}\
  \bibnamefont {Leidemann}}, \ and\ \bibinfo {author} {\bibfnamefont
  {Giuseppina}\ \bibnamefont {Orlandini}},\ }\bibfield  {title} {\enquote
  {\bibinfo {title} {{Response functions from integral transforms with a
  Lorentz kernel}},}\ }\href {\doibase 10.1016/0370-2693(94)91355-2} {\bibfield
   {journal} {\bibinfo  {journal} {Phys. Lett. B}\ }\textbf {\bibinfo {volume}
  {338}},\ \bibinfo {pages} {130--133} (\bibinfo {year} {1994})},\ \Eprint
  {http://arxiv.org/abs/nucl-th/9409004} {arXiv:nucl-th/9409004} \BibitemShut
  {NoStop}%
\bibitem [{\citenamefont {Raghavan}\ \emph {et~al.}(2021)\citenamefont
  {Raghavan}, \citenamefont {Balaprakash}, \citenamefont {Lovato},
  \citenamefont {Rocco},\ and\ \citenamefont {Wild}}]{Raghavan:2020bze}%
  \BibitemOpen
  \bibfield  {author} {\bibinfo {author} {\bibfnamefont {Krishnan}\
  \bibnamefont {Raghavan}}, \bibinfo {author} {\bibfnamefont {Prasanna}\
  \bibnamefont {Balaprakash}}, \bibinfo {author} {\bibfnamefont {Alessandro}\
  \bibnamefont {Lovato}}, \bibinfo {author} {\bibfnamefont {Noemi}\
  \bibnamefont {Rocco}}, \ and\ \bibinfo {author} {\bibfnamefont {Stefan~M.}\
  \bibnamefont {Wild}},\ }\bibfield  {title} {\enquote {\bibinfo {title}
  {{Machine learning-based inversion of nuclear responses}},}\ }\href {\doibase
  10.1103/PhysRevC.103.035502} {\bibfield  {journal} {\bibinfo  {journal}
  {Phys. Rev. C}\ }\textbf {\bibinfo {volume} {103}},\ \bibinfo {pages}
  {035502} (\bibinfo {year} {2021})},\ \Eprint
  {http://arxiv.org/abs/2010.12703} {arXiv:2010.12703 [nucl-th]} \BibitemShut
  {NoStop}%
\bibitem [{\citenamefont {Ekstr\"om}\ \emph {et~al.}(2015)\citenamefont
  {Ekstr\"om}, \citenamefont {Jansen}, \citenamefont {Wendt}, \citenamefont
  {Hagen}, \citenamefont {Papenbrock}, \citenamefont {Carlsson}, \citenamefont
  {Forss\'en}, \citenamefont {Hjorth-Jensen}, \citenamefont {Navr\'atil},\ and\
  \citenamefont {Nazarewicz}}]{Ekstrom:2015rta}%
  \BibitemOpen
  \bibfield  {author} {\bibinfo {author} {\bibfnamefont {A.}~\bibnamefont
  {Ekstr\"om}}, \bibinfo {author} {\bibfnamefont {G.~R.}\ \bibnamefont
  {Jansen}}, \bibinfo {author} {\bibfnamefont {K.~A.}\ \bibnamefont {Wendt}},
  \bibinfo {author} {\bibfnamefont {G.}~\bibnamefont {Hagen}}, \bibinfo
  {author} {\bibfnamefont {T.}~\bibnamefont {Papenbrock}}, \bibinfo {author}
  {\bibfnamefont {B.~D.}\ \bibnamefont {Carlsson}}, \bibinfo {author}
  {\bibfnamefont {C.}~\bibnamefont {Forss\'en}}, \bibinfo {author}
  {\bibfnamefont {M.}~\bibnamefont {Hjorth-Jensen}}, \bibinfo {author}
  {\bibfnamefont {P.}~\bibnamefont {Navr\'atil}}, \ and\ \bibinfo {author}
  {\bibfnamefont {W.}~\bibnamefont {Nazarewicz}},\ }\bibfield  {title}
  {\enquote {\bibinfo {title} {{Accurate nuclear radii and binding energies
  from a chiral interaction}},}\ }\href {\doibase 10.1103/PhysRevC.91.051301}
  {\bibfield  {journal} {\bibinfo  {journal} {Phys. Rev. C}\ }\textbf {\bibinfo
  {volume} {91}},\ \bibinfo {pages} {051301} (\bibinfo {year} {2015})},\
  \Eprint {http://arxiv.org/abs/1502.04682} {arXiv:1502.04682 [nucl-th]}
  \BibitemShut {NoStop}%
\bibitem [{\citenamefont {Epelbaum}\ \emph {et~al.}(2009)\citenamefont
  {Epelbaum}, \citenamefont {Hammer},\ and\ \citenamefont
  {Meissner}}]{Epelbaum:2008ga}%
  \BibitemOpen
  \bibfield  {author} {\bibinfo {author} {\bibfnamefont {Evgeny}\ \bibnamefont
  {Epelbaum}}, \bibinfo {author} {\bibfnamefont {Hans-Werner}\ \bibnamefont
  {Hammer}}, \ and\ \bibinfo {author} {\bibfnamefont {Ulf-G.}\ \bibnamefont
  {Meissner}},\ }\bibfield  {title} {\enquote {\bibinfo {title} {{Modern Theory
  of Nuclear Forces}},}\ }\href {\doibase 10.1103/RevModPhys.81.1773}
  {\bibfield  {journal} {\bibinfo  {journal} {Rev. Mod. Phys.}\ }\textbf
  {\bibinfo {volume} {81}},\ \bibinfo {pages} {1773--1825} (\bibinfo {year}
  {2009})},\ \Eprint {http://arxiv.org/abs/0811.1338} {arXiv:0811.1338
  [nucl-th]} \BibitemShut {NoStop}%
\bibitem [{\citenamefont {Hagen}\ \emph {et~al.}(2007)\citenamefont {Hagen},
  \citenamefont {Papenbrock}, \citenamefont {Dean}, \citenamefont {Schwenk},
  \citenamefont {Nogga}, \citenamefont {Wloch},\ and\ \citenamefont
  {Piecuch}}]{Hagen:2007ew}%
  \BibitemOpen
  \bibfield  {author} {\bibinfo {author} {\bibfnamefont {G.}~\bibnamefont
  {Hagen}}, \bibinfo {author} {\bibfnamefont {T.}~\bibnamefont {Papenbrock}},
  \bibinfo {author} {\bibfnamefont {D.~J.}\ \bibnamefont {Dean}}, \bibinfo
  {author} {\bibfnamefont {A.}~\bibnamefont {Schwenk}}, \bibinfo {author}
  {\bibfnamefont {A.}~\bibnamefont {Nogga}}, \bibinfo {author} {\bibfnamefont
  {M.}~\bibnamefont {Wloch}}, \ and\ \bibinfo {author} {\bibfnamefont
  {P.}~\bibnamefont {Piecuch}},\ }\bibfield  {title} {\enquote {\bibinfo
  {title} {{Coupled-cluster theory for three-body Hamiltonians}},}\ }\href
  {\doibase 10.1103/PhysRevC.76.034302} {\bibfield  {journal} {\bibinfo
  {journal} {Phys. Rev. C}\ }\textbf {\bibinfo {volume} {76}},\ \bibinfo
  {pages} {034302} (\bibinfo {year} {2007})},\ \Eprint
  {http://arxiv.org/abs/0704.2854} {arXiv:0704.2854 [nucl-th]} \BibitemShut
  {NoStop}%
\bibitem [{\citenamefont {Roth}\ \emph {et~al.}(2012)\citenamefont {Roth},
  \citenamefont {Binder}, \citenamefont {Vobig}, \citenamefont {Calci},
  \citenamefont {Langhammer},\ and\ \citenamefont {Navratil}}]{Roth:2011vt}%
  \BibitemOpen
  \bibfield  {author} {\bibinfo {author} {\bibfnamefont {Robert}\ \bibnamefont
  {Roth}}, \bibinfo {author} {\bibfnamefont {Sven}\ \bibnamefont {Binder}},
  \bibinfo {author} {\bibfnamefont {Klaus}\ \bibnamefont {Vobig}}, \bibinfo
  {author} {\bibfnamefont {Angelo}\ \bibnamefont {Calci}}, \bibinfo {author}
  {\bibfnamefont {Joachim}\ \bibnamefont {Langhammer}}, \ and\ \bibinfo
  {author} {\bibfnamefont {Petr}\ \bibnamefont {Navratil}},\ }\bibfield
  {title} {\enquote {\bibinfo {title} {{Ab Initio Calculations of Medium-Mass
  Nuclei with Normal-Ordered Chiral NN+3N Interactions}},}\ }\href {\doibase
  10.1103/PhysRevLett.109.052501} {\bibfield  {journal} {\bibinfo  {journal}
  {Phys. Rev. Lett.}\ }\textbf {\bibinfo {volume} {109}},\ \bibinfo {pages}
  {052501} (\bibinfo {year} {2012})},\ \Eprint {http://arxiv.org/abs/1112.0287}
  {arXiv:1112.0287 [nucl-th]} \BibitemShut {NoStop}%
\bibitem [{\citenamefont {Gu}\ \emph {et~al.}(2023)\citenamefont {Gu},
  \citenamefont {Sun}, \citenamefont {Hagen},\ and\ \citenamefont
  {Papenbrock}}]{Gu:2023aoc}%
  \BibitemOpen
  \bibfield  {author} {\bibinfo {author} {\bibfnamefont {Chenyi}\ \bibnamefont
  {Gu}}, \bibinfo {author} {\bibfnamefont {Z.~H.}\ \bibnamefont {Sun}},
  \bibinfo {author} {\bibfnamefont {G.}~\bibnamefont {Hagen}}, \ and\ \bibinfo
  {author} {\bibfnamefont {T.}~\bibnamefont {Papenbrock}},\ }\bibfield  {title}
  {\enquote {\bibinfo {title} {{Entanglement entropy of nuclear systems}},}\
  }\href@noop {} {\  (\bibinfo {year} {2023})},\ \Eprint
  {http://arxiv.org/abs/2303.04799} {arXiv:2303.04799 [nucl-th]} \BibitemShut
  {NoStop}%
\bibitem [{\citenamefont {Hagen}\ \emph {et~al.}(2010)\citenamefont {Hagen},
  \citenamefont {Papenbrock}, \citenamefont {Dean},\ and\ \citenamefont
  {Hjorth-Jensen}}]{Hagen:2010gd}%
  \BibitemOpen
  \bibfield  {author} {\bibinfo {author} {\bibfnamefont {G.}~\bibnamefont
  {Hagen}}, \bibinfo {author} {\bibfnamefont {T.}~\bibnamefont {Papenbrock}},
  \bibinfo {author} {\bibfnamefont {D.~J.}\ \bibnamefont {Dean}}, \ and\
  \bibinfo {author} {\bibfnamefont {M.}~\bibnamefont {Hjorth-Jensen}},\
  }\bibfield  {title} {\enquote {\bibinfo {title} {{Ab initio coupled-cluster
  approach to nuclear structure with modern nucleon-nucleon interactions}},}\
  }\href {\doibase 10.1103/PhysRevC.82.034330} {\bibfield  {journal} {\bibinfo
  {journal} {Phys. Rev. C}\ }\textbf {\bibinfo {volume} {82}},\ \bibinfo
  {pages} {034330} (\bibinfo {year} {2010})},\ \Eprint
  {http://arxiv.org/abs/1005.2627} {arXiv:1005.2627 [nucl-th]} \BibitemShut
  {NoStop}%
\bibitem [{\citenamefont {O'Connell}\ \emph {et~al.}(1987)\citenamefont
  {O'Connell} \emph {et~al.}}]{OConnell:1987ag}%
  \BibitemOpen
  \bibfield  {author} {\bibinfo {author} {\bibfnamefont {J.~S.}\ \bibnamefont
  {O'Connell}} \emph {et~al.},\ }\bibfield  {title} {\enquote {\bibinfo {title}
  {Electromagnetic excitation of the delta resonance in nuclei},}\ }\href@noop
  {} {\bibfield  {journal} {\bibinfo  {journal} {Phys. Rev.}\ }\textbf
  {\bibinfo {volume} {C35}},\ \bibinfo {pages} {1063} (\bibinfo {year}
  {1987})}\BibitemShut {NoStop}%
\bibitem [{\citenamefont {Anghinolfi}\ \emph {et~al.}(1996)\citenamefont
  {Anghinolfi} \emph {et~al.}}]{Anghinolfi:1996vm}%
  \BibitemOpen
  \bibfield  {author} {\bibinfo {author} {\bibfnamefont {M.}~\bibnamefont
  {Anghinolfi}} \emph {et~al.},\ }\bibfield  {title} {\enquote {\bibinfo
  {title} {Quasi-elastic and inelastic inclusive electron scattering from an
  oxygen jet target},}\ }\href@noop {} {\bibfield  {journal} {\bibinfo
  {journal} {Nucl. Phys.}\ }\textbf {\bibinfo {volume} {A602}},\ \bibinfo
  {pages} {405--422} (\bibinfo {year} {1996})},\ \Eprint
  {http://arxiv.org/abs/nucl-th/9603001} {nucl-th/9603001} \BibitemShut
  {NoStop}%
\bibitem [{\citenamefont {Abe}\ \emph {et~al.}(2020)\citenamefont {Abe} \emph
  {et~al.}}]{T2K:2020jav}%
  \BibitemOpen
  \bibfield  {author} {\bibinfo {author} {\bibfnamefont {K.}~\bibnamefont
  {Abe}} \emph {et~al.} (\bibinfo {collaboration} {T2K}),\ }\bibfield  {title}
  {\enquote {\bibinfo {title} {{Simultaneous measurement of the muon neutrino
  charged-current cross section on oxygen and carbon without pions in the final
  state at T2K}},}\ }\href {\doibase 10.1103/PhysRevD.101.112004} {\bibfield
  {journal} {\bibinfo  {journal} {Phys. Rev. D}\ }\textbf {\bibinfo {volume}
  {101}},\ \bibinfo {pages} {112004} (\bibinfo {year} {2020})},\ \Eprint
  {http://arxiv.org/abs/2004.05434} {arXiv:2004.05434 [hep-ex]} \BibitemShut
  {NoStop}%
\bibitem [{\citenamefont {Dytman}\ \emph {et~al.}(2021)\citenamefont {Dytman},
  \citenamefont {Hayato}, \citenamefont {Raboanary}, \citenamefont {Sobczyk},
  \citenamefont {Tena~Vidal},\ and\ \citenamefont
  {Vololoniaina}}]{Dytman:2021ohr}%
  \BibitemOpen
  \bibfield  {author} {\bibinfo {author} {\bibfnamefont {Steven}\ \bibnamefont
  {Dytman}}, \bibinfo {author} {\bibfnamefont {Yoshinari}\ \bibnamefont
  {Hayato}}, \bibinfo {author} {\bibfnamefont {Roland}\ \bibnamefont
  {Raboanary}}, \bibinfo {author} {\bibfnamefont {J.~T.}\ \bibnamefont
  {Sobczyk}}, \bibinfo {author} {\bibfnamefont {Julia}\ \bibnamefont
  {Tena~Vidal}}, \ and\ \bibinfo {author} {\bibfnamefont {Narisoa}\
  \bibnamefont {Vololoniaina}},\ }\bibfield  {title} {\enquote {\bibinfo
  {title} {{Comparison of validation methods of simulations for final state
  interactions in hadron production experiments}},}\ }\href {\doibase
  10.1103/PhysRevD.104.053006} {\bibfield  {journal} {\bibinfo  {journal}
  {Phys. Rev. D}\ }\textbf {\bibinfo {volume} {104}},\ \bibinfo {pages}
  {053006} (\bibinfo {year} {2021})},\ \Eprint
  {http://arxiv.org/abs/2103.07535} {arXiv:2103.07535 [hep-ph]} \BibitemShut
  {NoStop}%
\bibitem [{\citenamefont {Pastore}\ \emph {et~al.}(2020)\citenamefont
  {Pastore}, \citenamefont {Carlson}, \citenamefont {Gandolfi}, \citenamefont
  {Schiavilla},\ and\ \citenamefont {Wiringa}}]{Pastore:2019urn}%
  \BibitemOpen
  \bibfield  {author} {\bibinfo {author} {\bibfnamefont {Saori}\ \bibnamefont
  {Pastore}}, \bibinfo {author} {\bibfnamefont {Joseph}\ \bibnamefont
  {Carlson}}, \bibinfo {author} {\bibfnamefont {Stefano}\ \bibnamefont
  {Gandolfi}}, \bibinfo {author} {\bibfnamefont {Rocco}\ \bibnamefont
  {Schiavilla}}, \ and\ \bibinfo {author} {\bibfnamefont {Robert~B.}\
  \bibnamefont {Wiringa}},\ }\bibfield  {title} {\enquote {\bibinfo {title}
  {{Quasielastic lepton scattering and back-to-back nucleons in the short-time
  approximation}},}\ }\href {\doibase 10.1103/PhysRevC.101.044612} {\bibfield
  {journal} {\bibinfo  {journal} {Phys. Rev. C}\ }\textbf {\bibinfo {volume}
  {101}},\ \bibinfo {pages} {044612} (\bibinfo {year} {2020})},\ \Eprint
  {http://arxiv.org/abs/1909.06400} {arXiv:1909.06400 [nucl-th]} \BibitemShut
  {NoStop}%
\bibitem [{rep()}]{repo}%
  \BibitemOpen
  \href@noop {} {}\bibinfo {note}
  {\url{https://github.com/jesobczyk/spectral-function}}\BibitemShut {NoStop}%
\end{thebibliography}%

\end{document}